\documentclass[apjl,twocolumn]{emulateapj_mod}
\usepackage{epsfig,apjfonts,mathptmx}

\def\gtsima{$\; \buildrel > \over \sim \;$}
\def\ltsima{$\; \buildrel < \over \sim \;$}
\def\prosima{$\; \buildrel \propto \over \sim \;$}
\def\gsim{\lower.5ex\hbox{\gtsima}}
\def\lsim{\lower.5ex\hbox{\ltsima}}
\def\simgt{\lower.5ex\hbox{\gtsima}}
\def\simlt{\lower.5ex\hbox{\ltsima}}
\def\simpr{\lower.5ex\hbox{\prosima}}

\def\h1{$h^{-1}$}
\def\eeq{\end{equation}}
\def\beq{\begin{equation}}
\def\24mu{24\,$\mu{\rm m}$}
\def\70mu{70\,$\mu{\rm m}$}
\def\8mu{8\,$\mu{\rm m}$}

\shorttitle{Two bright submillimeter galaxies in a $z=4.05$
proto-cluster in GOODS-North}

\shortauthors{E. Daddi et al.}

\begin{document}

\title{
Two bright submillimeter galaxies in a $\lowercase{z}=4.05$
proto-cluster in GOODS-North, and accurate radio-infrared photometric
redshifts
}

   \author{E. Daddi\altaffilmark{1},
           H. Dannerbauer\altaffilmark{2},
           D. Stern\altaffilmark{3},
           M. Dickinson\altaffilmark{4},
           G. Morrison\altaffilmark{5,6},
           D. Elbaz\altaffilmark{1},
	   M. Giavalisco\altaffilmark{7},
           C. Mancini\altaffilmark{1,8},
	   A. Pope\altaffilmark{4},
           H. Spinrad\altaffilmark{9}
           }

\altaffiltext{1}{
    CEA, Laboratoire AIM - CNRS - Universit\'e Paris Diderot,
    Irfu/SAp, Orme des Merisiers, F-91191 Gif-sur-Yvette, France
    [e-mail: {\em edaddi@cea.fr}]}
\altaffiltext{2}{Max-Planck-Institut f\"ur Astronomie, K\"onigstuhl 17, D-69117 Heidelberg, Germany}
\altaffiltext{3}{Jet Propulsion Laboratory, California Institute of Technology, Pasadena, CA 91109}
\altaffiltext{4}{National Optical Astronomy Observatory,  950 N. Cherry Ave., Tucson, AZ, 85719}
\altaffiltext{5}{Institute for Astronomy, University of Hawaii, Honolulu, HI 96822}
\altaffiltext{6}{Canada-France-Hawaii Telescope, Kamuela, HI 96743}
\altaffiltext{7}{University of Massachusetts, Department of Astronomy, Amherst, MA 01003}
\altaffiltext{8}{Dipartimento di Astronomia, Universit\'a di Firenze, Largo E. Fermi 5, 50100 Firenze, Italia}
\altaffiltext{9}{Department of Astronomy, University of California at Berkeley, Mail Code 3411, Berkeley, CA 94720}

\begin{abstract}

We present the serendipitous discovery of molecular gas CO emission
lines with the IRAM Plateau de Bure interferometer coincident with
two luminous submillimeter galaxies (SMGs) in the Great Observatories
Origins Deep Survey North field (GOODS-N). The identification of
the millimeter emission lines as CO[4-3] at $z=4.05$ is based on
the optical and near-IR photometric redshifts, radio-infrared
photometric redshifts and Keck+DEIMOS optical spectroscopy.  These
two galaxies include the brightest submillimeter source in the field
(GN20; $S_{\rm 850\mu m}~=20.3$mJy, $z_{\rm CO}=4.055\pm0.001$)
and its companion (GN20.2; $S_{\rm 850\mu m}~=9.9$mJy, $z_{\rm
CO}=4.051\pm0.003$).  These are among the most distant
submillimeter-selected galaxies reliably identified through CO
emission and also some of the most luminous known.  
GN20.2 has a possible additional counterpart 
and a 
luminous AGN inside its primary counterpart revealed in the radio.
Continuum emission of
0.3mJy at 3.3mm (0.65mm in the rest frame) is detected at $5\sigma$
for GN20, the first dust continuum detection in an SMG at such long
wavelength, unveiling a spectral energy
distribution that is similar to local ultra luminous infrared
galaxies.  In terms of CO to bolometric luminosities, stellar mass
and star formation rates (SFRs), these newly discovered $z>4$ SMGs
are similar to $z\sim2-3$ SMGs studied to date.  
These $z\sim4$ SMGs have much higher specific
SFRs than typical B-band dropout Lyman break galaxies at the same redshift. The stellar
mass-SFR correlation for normal galaxies does not seem to evolve much further, between $z \sim 2$ and $z \sim
4$.  A significant $z=4.05$
spectroscopic redshift spike is observed in GOODS-N, and a strong spatial
overdensity of $B$-band dropouts and IRAC selected $z>3.5$ galaxies appears to be centered
on the GN20 and GN20.2 galaxies.
This suggests a proto-cluster structure with total mass $\sim
10^{14}M_{\odot}$.  
Using photometry at mid-IR (24$\mu$m), submm (850$\mu$m) and radio
(20cm) wavelengths, we show that reliable photometric redshifts
($\Delta z/(1+z)\sim0.1$) can be derived for SMGs over $1\simlt
z\simlt 4$.  This new photometric redshift technique has been used
to provide a first estimate of the space density of $3.5<z<6$
hyper-luminous starburst galaxies, and to show that they both
contribute substantially to the SFR density at early epochs and
that they can account for the presence of old galaxies at $z\sim2-3$.
Many of these high-redshift starbursts will be within reach of {\it
Herschel}.  We find that the criterion $S_{\rm 1.4~GHz}
\simgt S_{\rm 24\mu m}$, coupled to optical, near-IR and mid-IR
photometry, can be used to select $z>3.5$ starbursts, regardless of their
submm/mm emission.

\end{abstract}


\section{Introduction}

%The peak of the bolometric emission from dusty star forming galaxies
%falls in the 50-200$\mu$m wavelength range.  For this reasons,
Submillimeter and millimeter observations of galaxies are
subject to strong negative K-corrections over a very broad range
of redshifts, and can reach nearly constant bolometric luminosity
limits (for a given flux density limit) at $1 < z < 7$ (Blain et
al. 1993).  This provides a powerful method for detecting and
studying starburst galaxies in the distant universe.   Efforts to
identify counterparts of bright ($S_{\rm 850\mu m} > 5$~mJy)
submm-selected galaxies (SMGs; see Blain et al.\ 2002 for a review)
based on radio detections have recently suggested that the typical
redshift of these sources is $z\sim2$ (Chapman et al. 2003; 2005),
with few objects at $z\sim3$ and virtually none at $z>4$. This
result suggests a very strong decline in the space density of the
highest luminosity, most highly star forming galaxies, in the distant
universe.

SMGs are outliers in the stellar mass-star formation rate relation
in the $z\sim2$ universe (Daddi et al. 2007a; Takagi et al. 2008),
have gas kinematic signatures suggesting ongoing mergers (Tacconi
et al. 2006; 2008), have very compact sizes reminiscent of massive,
evolved galaxies at $z\sim1.5-3$ (Daddi et al. 2005a; Trujillo et
al. 2006; Zirm et al. 2007; Tacconi et al. 2006; 2008; Cimatti et
al. 2008) and have high star formation efficiencies compared to
typical galaxies of similar mass ($M\sim10^{10} - 10^{11}M_\odot$; Daddi et al. 2008; see also
Bouche et al. 2007). Their space density evolution
is a crucial issue for understanding the formation of massive
galaxies, as the SMGs are thought to represent massive galaxy mergers
which will rapidly evolve into passive systems (Daddi et al. 2007ab;
Tacconi et al. 2008; Cimatti et al. 2008).  Therefore, searching
for the very high redshift tail of SMGs holds clues into understanding
early type galaxy formation and characterizing the distribution of
their formation redshifts, both crucial issues for galaxy formation
models in a $\Lambda$CDM universe.

The strong decline of SMG counts at redshifts higher than $z\sim2$
might signal that $z\sim2$ is indeed the main formation epoch of
massive early-type galaxy systems (Daddi et al. 2004ab), but it can
be affected by selection biases, primarily the requirement of a
detection in the radio, which suffers from positive K-correction.
On the other hand, the existence of a population of passive systems
at $z\sim1.5-2.5$ (Spinrad et al. 1997; Cimatti et al. 2004;
McCarthy et al. 2004; Daddi et al. 2005a; Saracco et al.  2005;
Kriek et al. 2008) with large masses ($0.5-5\times10^{11}M_\odot$),
a space density of order of $10^{-4}$~Mpc$^{-3}$ ($\sim10-30$\%
of the correspondent one at $z=0$; Daddi et al. 2005a; Labb\'e et
al. 2005; Kong et al. 2006) and estimated formation redshifts for
their stars of $z\simgt3$ or even much higher (Daddi et al. 2005a;
Labb\`e et al. 2005; Maraston et al. 2006; Longhetti et al. 2007;
Cimatti et al. 2008), seems to require in turn the existence of a
substantial population of vigorous starburst galaxies at $z>3-4$,
which should be detectable from their strong far-infrared (FIR)
emission. For example, if this population of $z\sim1.5-3$ early
type galaxies were mainly formed during 100~Myr long bursts within
$3<z<6$, one would expect to find $\sim10^{-5}$~Mpc$^{-3}$ galaxies
with SFR$\simgt 1000 M_\odot$~yr$^{-1}$ in the same redshift range,
which would be detectable as SMGs and with a sky density of
$\approx200$~deg$^{-2}$.

Evidence has been mounting in the past years that such objects
actually exist.  For example,  a well known case is that of HDF850.1
(Dunlop et al. 2004), an SMG in the {\it Hubble} Deep Field that
is thought to be at very high redshift, lying behind a bright
foreground galaxy.  In addition, although largely lacking spectroscopic
identifications, the survey of Dannerbauer et al. (2002; 2004)
presented evidence that a high fraction of sources selected with the
1.2mm bolometer MAMBO at the IRAM 30~m telescope (Kreysa et al.
1998) are at high redshifts, given their very faint $K$-band and
radio detections or upper limits.  Similarly, Younger et al. (2007)
recently reported on seven sources selected with the AzTEC camera at
1.1mm (Wilson et al. 2008) with flux densities above 5~mJy in the
COSMOS field (Scoville et al. 2007), of which five (70\%) are found
to be very faint at optical and radio wavelengths, suggesting
redshifts $z\simgt3-4$.  More recently, Wang et al. (2007; 2008)
and Dannerbauer et al. (2008) presented high resolution submm/mm
imaging of a bright SMG from GOODS-N called GN10\footnote{Here and
in the remaining of the paper we adopt the ``GN'' nomenclature for
GOODS-N SMGs taken from Pope et al. (2006).}, suggesting that this
source is at a much higher redshift than thought before, probably
$3.5<z<6$. Follow-up surveys of high redshift radio galaxies also
resulted in submm detections (Archibald et al. 2001; Greve et al.
2007).  Knudsen et al. (2006) proposed a $z=4.048$ identification
for SMM~J16359$+$66130. However, this source lacks both radio and
CO identifications, and to our knowledge the result was never
published in a refereed journal.  Recently, Capak et al. (2008)
also suggest a $z=4.547$ identification for an SMG discovered with
AzTEC and IRAM in the COSMOS field. This high redshift galaxy is
$\simgt 1''$ away from the radio and IRAC positions. Offsets of up
to $\sim1''$ are indeed sometimes found for SMG counterparts (Chapman
et al. 2005). In a few of such cases the identification
turned out to be wrong (see, e.g., Pope et al. 2008), however Schinnerer et al. (2008)
confirmed the $z=4.5$ redshift for this galaxy.

Most of the evidence for $z>4$ SMGs comes from survey in the millimeter regime.
At first glance, it is not too clear how to reconcile these results
with much of the work done on SCUBA surveys follow-up. The slight
difference in selection wavelength (850$\mu$m versus 1.1mm) is
likely not sufficient to justify the diverging results. A possible
explanation is the trend for brighter SMGs to be systematically at
higher redshifts, as claimed by Ivison et al. (2002) and Pope et
al. (2006; P06 hereinafter).  Difficulties in locating the counterparts
within the large submm/mm positional uncertainties, and the fact
that the most distant sources could be ultra-faint or completely
undetected in the optical, near-IR, mid-IR, and/or radio, imply
that it is likely to obtain incorrect identifications by choosing
the brightest galaxy within the error circle of the submm position.
Given that star formation proceeds through the consumption of the
molecular gas reservoirs, the detection of CO emission is still the
most secure means to measure the redshift of an
SMG.

If SMGs are the progenitors of very massive early type galaxies,
signaling perhaps their last and strongest starburst before the
onset of passive evolution, they should also be strongly clustered
and reside in overdense regions of the universe. Blain et al. (2004)
report evidence for significant excess of redshift pairs in SMG
samples, suggesting significant clustering in this population.
Stevens et al. (2003) find significant excess of SMGs in the close
surroundings of high redshift radio galaxy fields, evidence that
SMGs might trace proto-cluster environments and be related to the
formation of present-day cluster ellipticals.  However, this result
is not yet spectroscopically confirmed. More recently, Chapman et al. (2008a)
discuss a $z=1.99$ concentration of five SMGs in the  GOODS-N field that 
does not correspond to a similarly strong overdensity of UV selected
galaxies. They suggest that concentration of SMGs might correspond to the
cosmic structures with the strongest enhancement in the frequency of major mergers.

In this paper, we take advantage of Plateau de Bure Interferometer
(PdBI) observations at millimeter wavelengths in the GOODS-N area
that serendipitously revealed two strong emission lines, coincident
with the positions of GN20, the brightest SMG in the GOODS-N field
($S_{\rm 850 \mu m}=20.3$mJy), and GN20.2 ($S_{\rm 850 \mu m}=9.9$mJy).
Using multiwavelength photometric imaging data and Keck spectroscopy,
we identify the lines as CO[4-3] transitions at $z=4.055\pm0.001$
and $z=4.051\pm0.003$.  Thanks to the serendipitous detection of
CO lines in the PdBI field of view,
%\footnote{To our knowledge, these
%are the first serendipitous CO detections from the distant universe
%to date.}, 
this reverses the typical SMG counterpart identification
process in which a plausible optical/near-IR counterpart with a
measured spectroscopic redshift is only later confirmed through CO
observations.  In addition, we provide spectroscopic (CO) and
photometric evidence for a second plausible counterpart to GN20.2,
likely at the same redshift. This concentration of submm galaxies
appears to define a proto-cluster structure as traced by $B$-band
dropout Lyman break galaxies and massive IRAC selected $z>3.5$
galaxies.  The newly measured $z>4$ CO redshifts for SMGs have been
exploited to calibrate and test the use of radio-IR photometric
redshifts for the SMG population, and to obtain a new estimate of
the abundance of very high redshift SMGs.

\begin{figure*}
\centering
\includegraphics[width=16.0cm,angle=0]{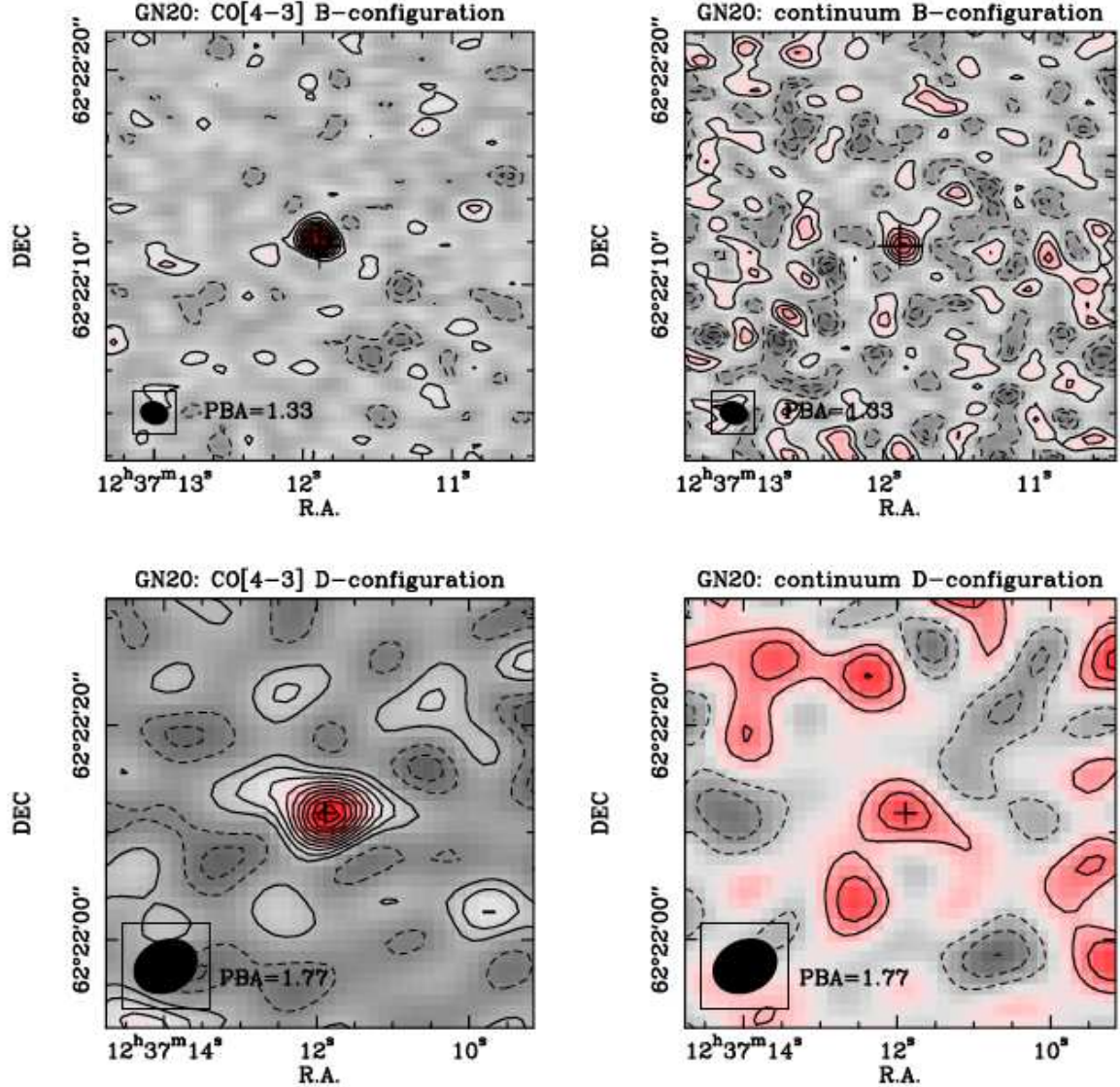}
\caption{
Velocity averaged spatial maps at 91.4~GHz
centered on the serendipitously detected emission from GN20 (the $2''$ wide cross refers to the VLA 1.4~GHz position). 
The two top panels show B-configuration 1.3$''$ beam maps (20$''$ field of view) while
the bottom panels show the D-configuration 5.5$''$ beam ones (40$''$ field of view). 
Red colors correspond to positive fluxes and gray colors to negative fluxes. Contours level 
are shown in steps of 1$\sigma$ with positive (negative) contours shown as solid (dashed)
lines. The CO[4-3] 
maps are averaged over 750 km s$^{-1}$ 
corresponding to the observed range
of CO emission (see Fig.~\ref{fig:1D}) and were cleaned. The continuum maps show the average 
of all channels outside the line range (for a 2250 km/s total bandpass).
Notice that GN20 is observed off from the phase center with primary beam attenuation
(PBA, labeled in each image) of 1.33 and 1.77 at its position in the B- and D-configuration observations (these
maps were not corrected for attenuation in order to preserve the flatness of the noise). Given the relatively low 
S/N ratios, the continuum maps were not cleaned as the side lobes are below the noise level.
}
\label{fig:2Da}
\end{figure*}

The paper is organized as follows. In Sect.~\ref{sec:pdbi} we
describe the PdBI CO line detections, and determine their redshifts
in Sect.~\ref{sec:redshift}.  The physical properties of the sources,
in terms of molecular gas, SFRs and masses, are discussed in
Sect.~\ref{sec:CO}. In Sect.\ref{sec:cluster} we show that these
objects lie at the center of a proto-cluster environment. We describe
a method to obtain accurate photometric redshifts for SMGs in
Sect.\ref{sec:phoz}, and discuss the properties and present a new
selection technique for identifying high redshift starbursts in
Sect.\ref{sec:z4}.  Throughout the paper magnitudes are expressed
in the AB scale unless stated otherwise, and a standard cosmology
is adopted with $\Omega_\Lambda, \Omega_M = 0.76, 0.24$, and $h =
H_0/$(100 {\rm km s}$^{-1}$ {\rm Mpc}$^{-1}$)$=0.73$.

\begin{figure*}
\centering
\includegraphics[width=16.0cm,angle=0]{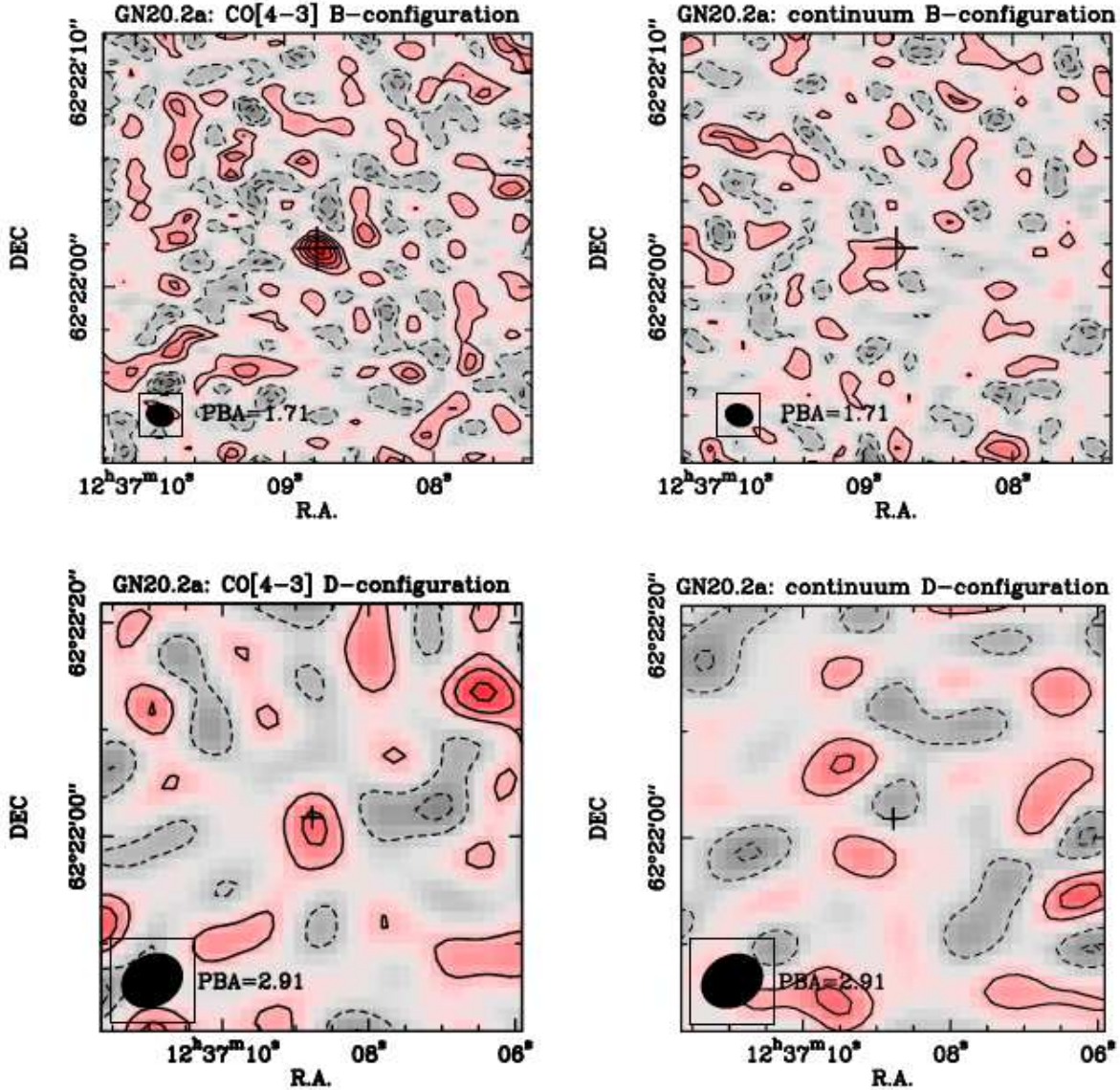}
\caption{The same of Fig.~\ref{fig:2Da} but for the galaxy GN20.2a (the 2$''$ wide
cross refers to the VLA 1.4~GHz position).
The CO[4-3] maps are averaged here over 1125 km s$^{-1}$ 
corresponding to the observed range
of CO emission for GN20.2a (see Fig.~\ref{fig:1D}). The continuum maps show the average 
of all channels outside the line range (for a 1875 km/s total bandpass).
GN20.2a is even more offset than GN20 from the phase center with primary beam attenuation
(PBA, labeled in each image) of 1.71 and 2.91 at its position in the B- and D-configuration observations.
Before creating these maps, the bright emission at the position of GN20 was fitted in the {\em uv}
data with a point source model and subtracted, in order to avoid contamination at the position of GN20.2
from the side     
lobes of the more luminous GN20 galaxy. These maps were not cleaned as the side lobes are below the noise level
in all cases. No correction  for PBA is applied to these maps, in order to preserve the flatness of the noise.
}
\label{fig:2Db}
\end{figure*}

\section{PdBI observations}
\label{sec:pdbi}

We have used the IRAM PdBI to map the CO[2-1] transition from the
$z=1.522$ galaxy BzK-21000 (J123710.60+622234.6)  redshifted to about 91.4~GHz. 
We obtained
7.5 hours of on-source integration in D-configuration (5.5$''$
synthesized beam) during April 2007 and 6.4~hours
on source follow-up observations were obtained in January 2008 with a 
B-configuration (1.3$''$ synthesized beam).
For both D- and B-configuration observations the PdBI
was used with all 6 antennas available (15 independent baselines).
The correlator consists of 8 independent units each of which was covering 320~MHz
(128 channels each with a width of 2.5~MHz) with a single polarization, covering a total bandwidth of about 1~GHz
with both polarizations. 
Observations were tuned at a central frequency of 91.375~GHz (for a total velocity bandpass of about 3000~km/s).
For calibration of the data we observed standard bandpass calibrators
(J0418$+$380, 3C273), phase/amplitude calibrators (J1044$+$719, J1150$+$497), and
flux calibrators (MWC~349, 3C273) to which we regularly switched during
the primary target observations.  We reduced the data with the GILDAS software packages CLIC
and MAP. The maps of the fields obtained using natural weights and using the full 1~GHz bandpass have noise
levels of about 67.9$\mu$Jy/beam and 51.2$\mu$Jy/beam for the D- and B-configuration data, respectively.

\begin{figure*}
\centering
\includegraphics[width=16.5cm,angle=0]{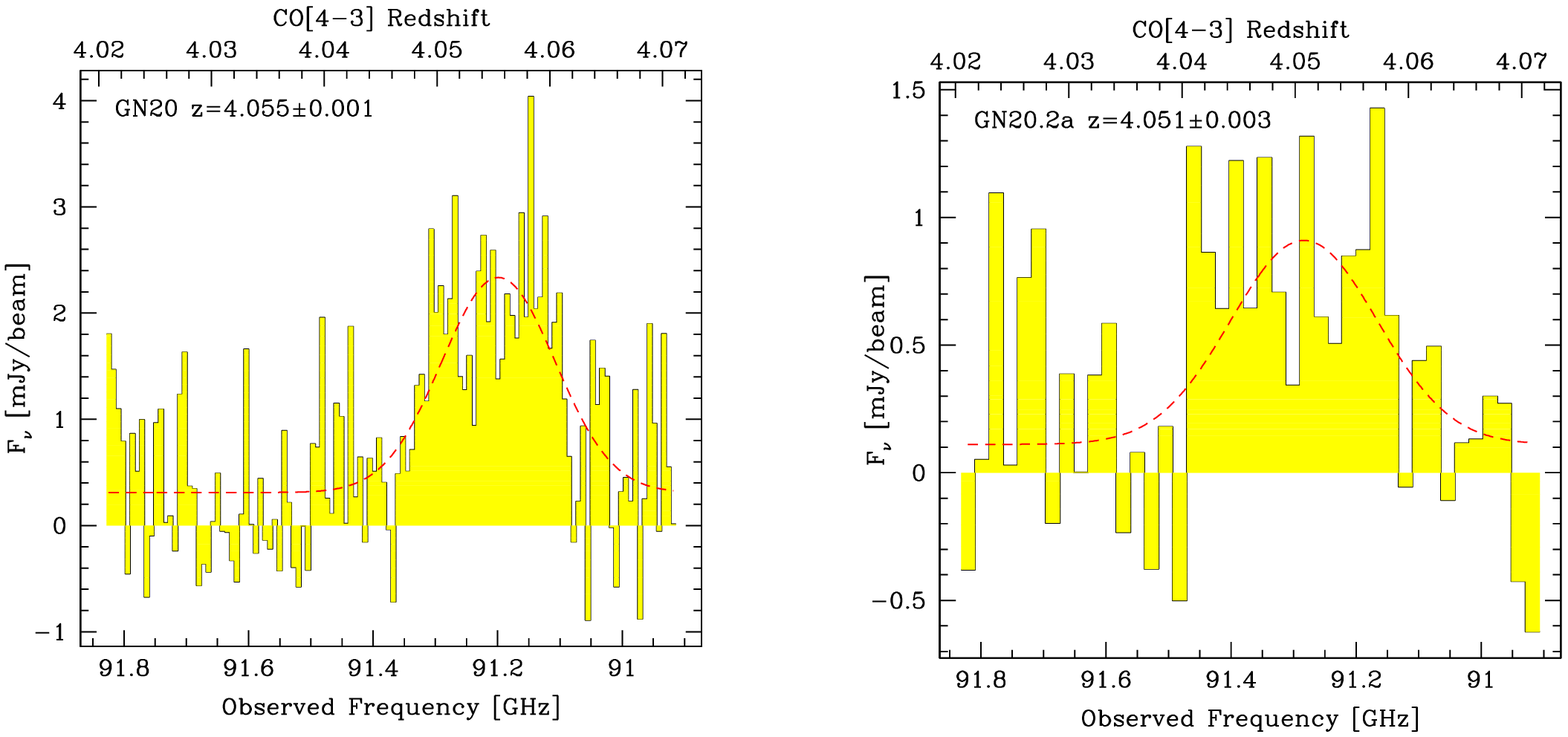}   
\caption{
One dimensional spectra of CO[4-3] from GN20 (left, sampled at 25~km
s$^{-1}$) and GN20.2a (right, sampled at 75~km s$^{-1}$).  The
flux densities have been corrected for primary beam attenuation.
The red curve shows the best fitting Gaussian to the measured signal,
allowing for the presence of faint underlying continuum. 
}
\label{fig:1D}
\end{figure*}

\begin{deluxetable*}{lcccc}%[h]
\tabletypesize{\scriptsize}
\tablecaption{Radio and CO coordinates of the sources}
\tablewidth{0pt}
\tablehead{
\colhead{ID} &
\colhead{RA$_{\rm CO}$(J2000)}&
\colhead{DEC$_{\rm CO}$(J2000)}&
\colhead{RA$_{\rm VLA}$(J2000)}&
\colhead{DEC$_{\rm VLA}$(J2000)}
}
\startdata
GN20 & 12:37:11.90 & 62:22:12.1 & 12:37:11.89 & 62:22:11.8 \\
GN20.2a & 12:37:08.77 & 62:22:01.7 & 12:37:08.78 & 62:22:01.8 \\
GN20.2b & -- & -- & 12:37:09.73 & 62:22:02.6 
\enddata
\tablecomments{The CO coordinates are from our 1.3$''$ beam, B-configuration observations of CO[4-3]. 
The VLA coordinates are from the 1.4~GHz radio continuum map of Morrison et al. (in preparation) 
with 1.7$''$ beam.
}
\label{tab:radec}
\end{deluxetable*}

Daddi et al. (2008) presented the CO[2-1] detection of the BzK galaxy with the D-configuration data.
In addition to the BzK galaxy, from the initial D-configuration observations we also found
a strong, serendipitous source in the field, located 22$''$ South and
8$''$ East of the primary target (Fig.~\ref{fig:2Da}).  Even when averaging all available
channels over the 3000~km s$^{-1}$ spectral range observed, and
after subtracting from the data the signal from the primary target,
the serendipitous source was securely detected, despite the large
primary beam attenuation at its distance from phase center (the
primary beam results in an approximately Gaussian response with
full width at half maximum [FWHM] of $55''.2$ at this frequency).
For this reason, for the follow-up observations in the more extended B-configuration
we offset the pointing by 10$''$ from the BzK
galaxy in order to reduce the primary beam attenuation at the
position of the serendipitous detection without compromising the
data quality at the position of the primary target. In addition to
confirming this serendipitous source, the new observations also
revealed a second source at high significance (Fig.~\ref{fig:2Db}
and \ref{fig:1D}), and a third tentative detection. The two secure
detections correspond well to the positions of GN20 and GN20.2 in
the sample of SMGs published in P06. In both cases, it is clear
that most of the detected signal comes from an emission line, most
likely from CO molecular gas\footnote{It is unlikely that the
emission line is not from CO, given that CO lines are the brightest
lines observed at millimeter wavelengths.}.  Measured quantities
are summarized in Tables~\ref{tab:radec}~and~\ref{tab:GN20} and derived
quantities (like luminosities and masses) are in Table~\ref{tab:GN20_est}.

\subsection{CO detection of GN20}

\begin{figure*}
\centering
\includegraphics[width=13cm,angle=0]{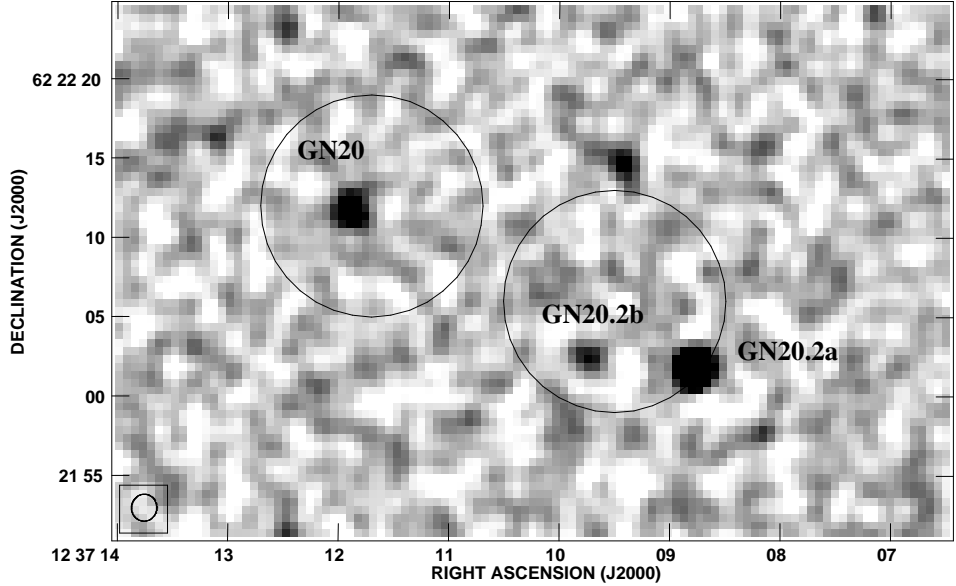}   
\caption{VLA 1.4~GHz radio continuum map of the GN20 (left) and GN20.2 (right) region.
The VLA observations were mostly taken in A-array and have a beam of 1.7$''$ FWHM
(Morrison et al., in preparation).
The 7$''$ radii circles are centered on the SCUBA positions
from P06. The radio sources within the 850$\mu$m beam are labeled. 
%The colorbar at the bottom
%shows the flux level in Jansky. 
The faint radio emission about 1$''$ North of the GN20.2 SCUBA beam
is real and corresponds to a galaxy with $z_{phot}=1.6$. From its mid-IR emission we infer
only moderate SFR levels that should not be substantially contributing
to the flux density observed at submm wavelength at the position of GN20.2.
}
\label{fig:vla}
\end{figure*}

GN20 is very securely detected in both D- and B-configuration observations
(Fig.~\ref{fig:2Da} left). Most of the signal appears to be due to a  CO emission
line. Averaging the signal from 150~km~s$^{-1}$ to 900~km~s$^{-1}$ (here and in the
following all velocities are defined with respect to the central
frequency of the observations at 91.375~GHz) results in a detection
with $S/N=10$ for the D-configuration data (Fig.~\ref{fig:2Da} left and Fig.~\ref{fig:1D} top).
The B~configuration observations within the same velocity range results in a detection
with $S/N=13$ and a position of $\alpha(J2000)= 12^h37^m11^s.90$,
$\delta(J2000)= 62^d22^m12^s.1$ (Fig.~\ref{fig:2Da} left).  This position is consistent with
the $7''$ submm error box of the SMG GN20 (P06), and is within
$0.5''$ of the accurate interferometric Iono et al. (2006)
Sub-Millimeter-Array (SMA) detection of this source and of the radio
counterpart in 20cm VLA data (Morrison et al., in preparation;
Fig.~\ref{fig:vla}). The PdBI position coincides well with a faint
galaxy that P06 already identified with the optical counterpart of
GN20 (Fig.\ref{fig:ACS}).  GN20 is the brightest SMG in the GOODS-N
area. The radio, CO, and SMA submm positions actually appear to be
slightly offset to the East of the source seen in the optical (by
$\sim0.5''$ or so), as also noted by Iono et al. (2006).  It is not
clear if this is due to some additional galactic structure being
present but extremely obscured, or just a random fluctuation across
the data sets.

The PdBI D- and B-configuration datasets were
independently fitted
with a point source 
and the resulting spectra were corrected for primary beam attenuation (PBA) and coadded
with appropriate weighting (the D-configuration observations are noisier on an absolute flux scale, 
also due to the larger correction for PBA).  We performed Gaussian fitting of the resulting spectrum (Fig.~\ref{fig:1D})
to determine the CO line properties, 
allowing for the presence of a faint underlying continuum.  The best fitting
Gaussian line is offset by 580$\pm40$~km~s$^{-1}$ relative to the
central frequency, and has an velocity integrated flux of {I$_{\rm
CO}=1.5\pm0.2$~Jy~km~s$^{-1}$}, with both errors given at the
$1\sigma$ level and following Avni (1976) for the case of two
parameters of interest (line velocity and integrated emission line
flux).  There is evidence for faint 3mm continuum as well, as shown
in Fig.~\ref{fig:2Da} (right panels) and Fig.~\ref{fig:1D}. To confirm this result, we averaged the signal
in UV space for velocities outside a 1200~km~s$^{-1}$ range centered
on the line (about 1.7$\times$FWHM), where negligible line contribution
should be present. The resulting continuum map has a faint source at a
position consistent within the errors with that of the emission line and
corresponding to a signal of 0.33~mJy, significant at the 5$\sigma$ level.  To our
knowledge, this is the first secure dust continuum detection of an
SMG in the 3mm band.

\subsection{CO detection of GN20.2a}

Averaging the velocity channels between -300~km~s$^{-1}$ and
825~km~s$^{-1}$ a source is detected in the PdBI map with $S/N=5.8$ at the position
of $\alpha(J2000)= 12^h37^m08^s.77$, $\delta(J2000)= 62^d22^m01^s.7$
in the B-configuration observations
(Fig.~\ref{fig:2Db} left and Fig.~\ref{fig:1D} right). Positive flux with $S/N$ just below 3
is observed  also in the D-configuration data at this position (Fig.~\ref{fig:2Db} left). We
extracted the spectra by fitting point sources in the D- and B-configuration datasets independently 
and coadded the spectra with weighting after correcting for primary beam attenuation. Averaging
the total resulting spectrum within  -300~km~s$^{-1}$ and
825~km~s$^{-1}$ we find a positive signal significant at the $7\sigma$ level.

\begin{figure*}
\centering
\includegraphics[width=16.0cm,angle=0]{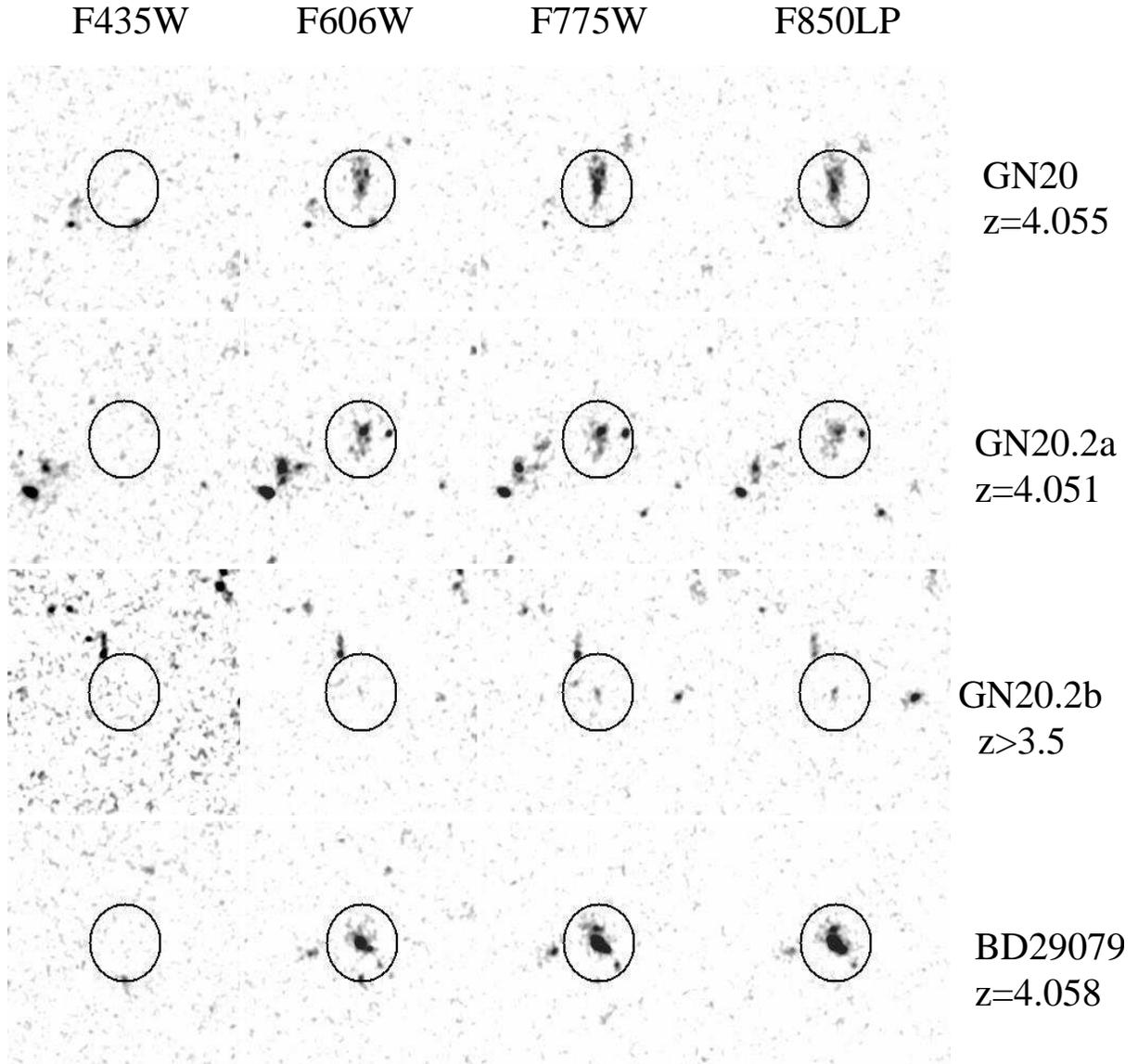}   
\caption{Multiband {\it HST}+ACS imaging of the optical counterparts of
GN20 (top), GN20.2a (second from top), GN20.2b (third from top) and
a B-dropout Lyman break galaxy in the area having also $z_{\rm spec}=4.058$
from Keck spectroscopy (the coordinates of this source,
named BD29079 in the GOODS-N B-dropout catalog, are
$\alpha(J2000)= 12^h37^m11^s.53$, $\delta(J2000)= 62^d21^m55^s.7$).
From left to right, as labeled, the data refer to the ACS filters
F435W, F606W, F775W and F850LP (sometimes referred to in the text also as
$B$-, $v$-, $i$- and $z$-band, respectively). The images have been smoothed with
a 3 pixel FWHM Gaussian. Circles around the sources have $1''$
radius and are simply centered on the identified ACS counterparts. 
}
\label{fig:ACS}
\end{figure*}

The position of this source is within 0.2$''$ of the relatively
bright radio source that P06 identified with the counterpart of the
SMG GN20.2, close but well distinguished from GN20 that is 
$24''$ to the NE. The position corresponds to a faint galaxy in the
deep {\it Hubble Space Telescope} ({\it HST}) images (Fig.~\ref{fig:ACS}).
Also in this case the signal appears to be mostly due to an emission
line. The best fitting Gaussian line is offset by 330$\pm150$~km~s$^{-1}$
relative to the central frequency, and has a velocity integrated
flux of {I$_{\rm CO}=0.9\pm0.3$~Jy~km~s$^{-1}$}. The largest
contribution to the flux error is the uncertainty in the continuum
level. Some positive continuum appears to be present, although the
$S/N$ is not yet sufficient for a reliable measurement (Tab.~\ref{tab:GN20}).
Accounting for the possible continuum emission and its uncertainty, the CO emission
line is still detected at the $6\sigma$ level.

\subsection{GN20.2b: a second counterpart to GN20.2 ?}
\label{sec:20p2b}

Inspection of the newly reprocessed 1.4~GHz radio map around the
position of GN20.2 (Fig.~\ref{fig:vla}; Morrison et al., in
preparation) shows an additional faint radio source with a flux
density of $32.2\mu$Jy (5$\sigma$) at $\alpha(J2000)=
12^h37^m09^s.73$, $\delta(J2000)= 62^d22^m02^s.6$ (radio frame),
3.4$''$ from the GN20.2 SCUBA position (P06) and 6.8$''$ from the
CO detected counterpart discussed above. We refer to the primary
counterpart of GN20.2 as GN20.2a, and this newly identified faint
radio source is referred to as GN20.2b.  GN20.2b is a plausible
additional counterpart to the submm emission.

Interestingly, positive signal is found in the PdBI map close to the
position of GN20.2b. 
The PdBI D- and B-configuration datasets were
independently fit
with a point source fixed at the VLA position of GN20.2b (a secure
position cannot be derived from the PdBI data alone, due to the low
$S/N$) and the resulting spectra were corrected for primary beam attenuation (PBA) and coadded
weighting according to the noise,
similarly to what was done for the other galaxies. 
Averaging over the whole 1~GHz bandwidth in the coadded spectrum
results in a $\sim3\sigma$ signal.
If we average  the spectrum between $\approx0$~km~s$^{-1}$ and
$\approx700$~km~s$^{-1}$, we find a $>4\sigma$ signal. Although longer
integrations at PdBI would be required for a more reliable conclusion,
the data are again consistent with CO emission, quite close in
frequency to the other securely detected lines. If real, that would correspond
to an integrated flux of {I$_{\rm CO}\approx0.45$~Jy~km~s$^{-1}$}.

\section{Redshift identifications}
\label{sec:redshift}

\begin{figure}
\centering
\includegraphics[width=8.8cm,angle=0]{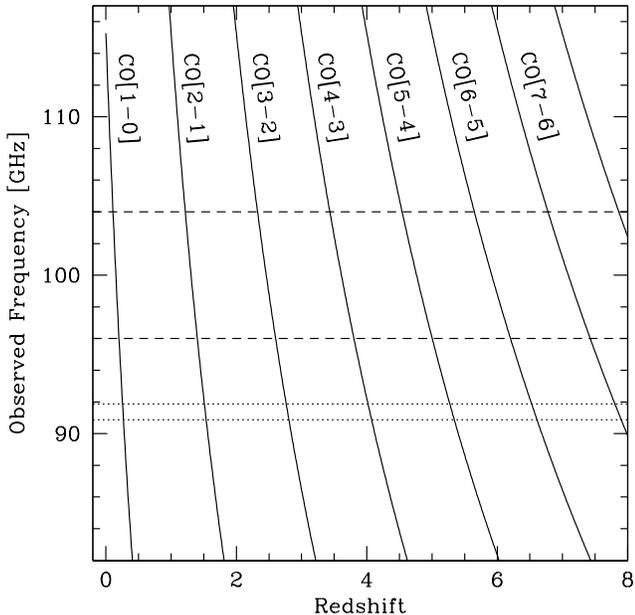}   
\caption{The frequency-redshift relation for CO emission lines in
the 3mm atmospheric window. The dotted range around 91.4~GHz
corresponds to the PdBI observations presented in this paper. The
dashed line centered at 100~GHz corresponds to an 8~GHz range, as
planned for future generations of correlators, like those that will
be used on ALMA.  }
\label{fig:ALMA}
\end{figure}

Depending on which transition we are actually observing, the CO
detection fixes the redshift of GN20 and GN20.2a to $1+z_{\rm
CO}\sim 1.264\times n$, where the observed transition is
$CO[(n)-(n-1)]$.  Possible redshifts are: 0.26, 1.53, 2.80, 4.05,
5.32, 6.58, etc. The most straightforward guess for the redshift would
be 1.53, corresponding to companion galaxies to BzK-21000, the
primary target of the PdBI observations. However, we show
in the following that the
correct identification for the lines is CO[4-3] at redshift
$z_{\rm CO}\sim4.05$, making these among the most distant CO detections
so far for SMGs. Although it might seem unlikely that we have
serendipitously detected CO emission lines from $z\simgt4$ galaxies,
we note that the 1~GHz spectral range of our data maps into a 
twice larger redshift range at $z\sim4$ versus $z\sim1.5$, which
in turn corresponds to a two times larger comoving volume.  Also,
at higher redshift the density of CO transitions per unit frequency
and at fixed observed frequency is higher, hence chance detections
are non-negligible once a luminous high redshift target is within
the field of view (Fig.~\ref{fig:ALMA}).

The following section discusses several independent redshift estimates
for these galaxies, e.g., based on optical photometric redshifts,
radio-IR photometric redshifts, and optical spectroscopy. This
comparison is interesting as the problem of unidentified CO line
detections will become a typical situation of the coming years with
new, wide-band instrumentation rapidly becoming available. Even
$8-20$~GHz receivers will often identify single CO lines from single
galaxies (Fig.~\ref{fig:ALMA}).  A direct assessment and inter
comparison of multiple redshift identification techniques is thus
interesting for this first test case of serendipitously detected
CO lines.

\begin{figure*}
\centering
\includegraphics[width=16.0cm,angle=0]{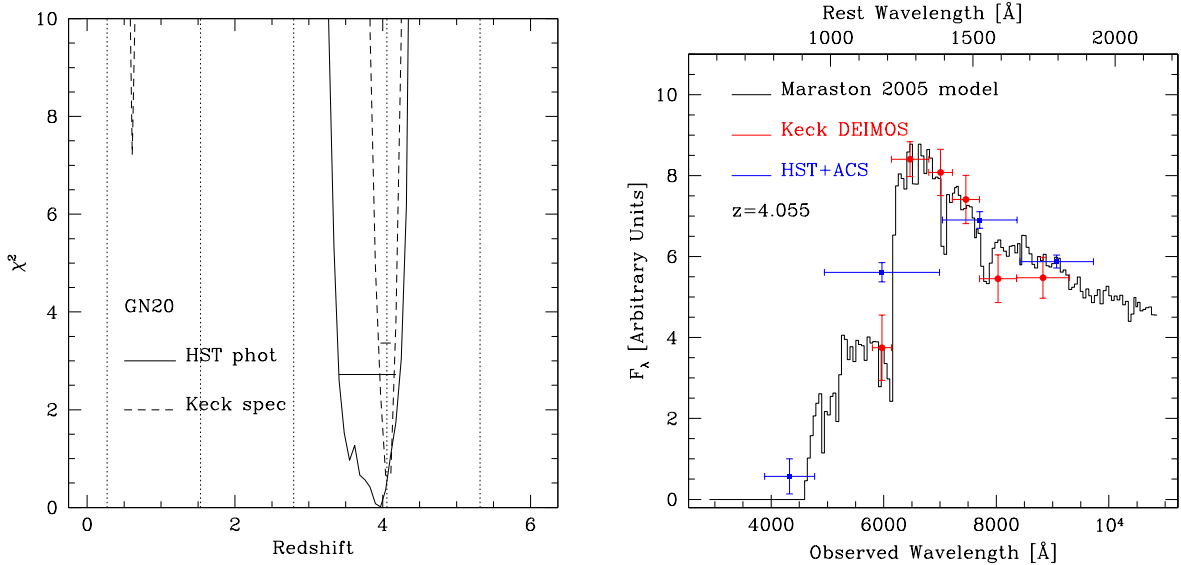}   
\caption{
Left panel: 
photometric redshifts from stellar emission for the optical
counterpart of GN20. The solid line shows the result of fitting the UV
rest frame continuum {\it HST} photometry described in Section~\ref{sec:chi2P}, 
while the dashed line refers to the fit of the binned Keck+DEIMOS spectroscopy
discussed in Section~\ref{sec:kspec}. The dotted horizontal lines show the permitted
redshifts, based on the detection of CO emission.
Right panel: the optical SED of GN20 as derived from {\it HST}+ACS imaging (blue points) and
Keck spectroscopy (after binning in wavelength; red points). A strong break at about 6150\AA\
is observed, corresponding to the Ly$\alpha$ forest break at $z=4.055$.
}
\label{fig:chi2P}
\end{figure*}

\begin{figure*}
\centering
\includegraphics[width=16.0cm,angle=0]{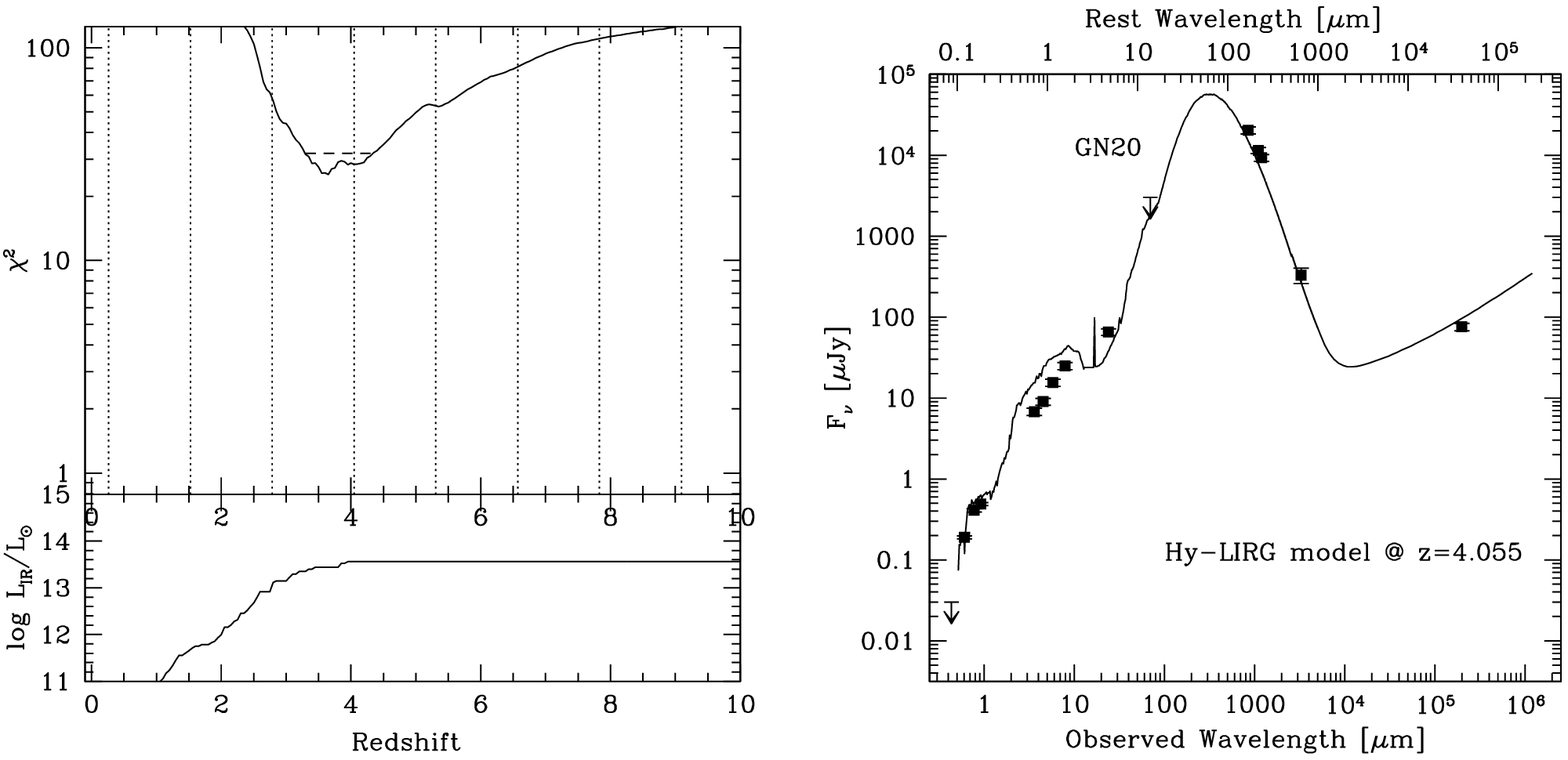}   
\caption{Left panel: $\chi^2$ versus redshift for the radio-IR
fitting of GN20, based on the 24$\mu$m, 850$\mu$m, 3mm
and 20cm observed flux densities and the CE01 library of template galaxies.
Right panel: the SED of GN20. The brightest CE01 model with $L_{\rm IR}=3.5\times10^{13}L_\odot$ is
overplotted for comparison, without any rescaling, redshifted to $z=4.055$. This is very close to
the best fitting photometric redshift solution at $z_{phot}=3.7$.
We show here also the 1.1mm and 1.2mm flux
density measurements recently published by Perera et al. (2008) and Greve et al. (2008). 
GN20 is undetected in the 70$\mu$m {\em Spitzer}+MIPS imaging by Frayer et al. (2006) 
and we show an upper limit of 3~mJy.
}
\label{fig:chi2F}
\end{figure*}

\subsection{GN20}
\label{sec:GN20}

\subsubsection{Photometric redshift from stellar emission}
\label{sec:chi2P}

Fig.~\ref{fig:ACS} shows {\it HST}+ACS imaging in the four bands
available in GOODS (Giavalisco et al. 2004a).  As already noted by
P06, the optical counterpart of GN20 is a B-dropout Lyman break
galaxy according to the definition of Giavalisco et al.  (2004b).
This definition returns galaxy samples with $z=3.78 \pm 0.34$,
favoring the identification of the PdBI line as CO[4-3] at $z=4.055$.
Circumstantial evidence for the likelihood of a redshift around
$z=4.05$ had been even earlier presumed from the presence of a
bright $B$-band dropout galaxy at a 16$''$ separation (Fig.~\ref{fig:ACS})
with fairly similar colors and a known spectroscopic redshift of
$z=4.058$ measured from Keck spectroscopy (Stern et al., in
preparation, we refer to this object as BD29079).  The IRAC colors of the GN20 $B$-band dropout counterpart
are very red, with the brightest IRAC band being the 8.0$\mu$m (see,
e.g., P06). If due to the emission of stars, this implies that the
1.6$\mu$m rest frame bump is beyond the 5.8$\mu$m channel, or
$z\simgt3.5$. However, the IRAC emission can also be red due to,
or affected by, the presence of obscured AGNs (e.g., Stern et al.
2005; Daddi et al. 2007b).

Stellar photometric redshifts were thus estimated using only the four
{\it HST} bands and running the {\em hyperz} code (Bolzonella et
al. 2002).  The Maraston (2005) models were used, with a variety
of star formation histories and allowing for reddening using the
Calzetti et al. (2000) law.  The near-IR bands are not used because
the available datasets, described, e.g., in Daddi et al. (2007a),
are not sufficiently deep for placing meaningful constraints on the
 spectral energy distribution
SED of this faint and relatively blue galaxy.  The distribution of
$\chi^2$ values versus redshift (Fig.\ref{fig:chi2P}) is consistent
with $3.4<z<4.2$ (95\% confidence range).  Fig.~\ref{fig:chi2P}
shows the observed SED and the best fitting model for
the case of $z=4.055$.

\begin{deluxetable*}{lcccccccccc}%[h]
%\tabletypesize{\scriptsize}
\tablecaption{Measured properties of GN20, GN20.2a and GN20.2b}
\tablewidth{0pt}
\tablehead{
\colhead{ID} &
\colhead{$z_{\rm CO}$}&
\colhead{$z_{\rm Keck}$}&
\colhead{$S_{\rm 850\mu m}$}&
\colhead{$S_{\rm 3.3mm}$}&
\colhead{$S_{\rm 1.4 GHz}$}&
\colhead{$S_{\rm 24\mu m}$}&
\colhead{$S_{\rm 8.0\mu m}$}&
\colhead{$S_{\rm 4.5\mu m}$}&
\colhead{$I_{\rm CO[4-3]}$}&
\colhead{FWHM}
\\
\colhead{}&
\colhead{}&
\colhead{}&
\colhead{(mJy)}&
\colhead{(mJy)}&
\colhead{($\mu$Jy)}&
\colhead{($\mu$Jy)}&
\colhead{($\mu$Jy)}&
\colhead{($\mu$Jy)}&
\colhead{(Jy km s$^{-1}$)}&
\colhead{(km~s$^{-1}$)}
\\
\colhead{(1)}&
\colhead{(2)}&
\colhead{(3)}&
\colhead{(4)}&
\colhead{(5)}&
\colhead{(6)}&
\colhead{(7)}&
\colhead{(8)}&
\colhead{(9)}&
\colhead{(10)}&
\colhead{(11)}
}
\startdata
GN20 & $4.055\pm0.001$ & 4.06$\pm$0.02 & 20.3$\pm$2.1$^{\dagger}$ &
0.33$\pm$0.07 & 75.8$\pm7.9$& 65.5$\pm$3.5 &  $26.1\pm2.6$ & $9.5\pm1.0$ &  1.5$\pm$0.2 & $710\pm120$ \\  
GN20.2a & $4.051\pm0.003$ &4.059 $\pm$0.007 &$\simlt$9.9$\pm$2.3$^{\dagger}$ &
$<0.20$ & 180.7$\pm$8.4$^{\dagger}$& $30.2\pm5.6$ & $9.8\pm1.0$ &  $4.1\pm0.4$ &  $0.9\pm0.3$ & $1100\pm400$ \\  
GN20.2b & --  & -- & $<$9.9$\pm$2.3$^{\dagger}$ & $<0.18$ &
32.2$\pm$6.5 & $12.0\pm4.3$ & $15.4\pm1.5$ &  $10.1\pm1.0$ & -- & -- 
\enddata
\tablecomments{Col.~(1): Name of the object. Col.~(2): Redshift derived
from the CO[4-3] observations. 
Col.~(3): Redshift derived from the Keck optical spectroscopy.
Col.~(4): SCUBA 850~$\mu$m flux density from P06$^{\dagger}$.
Col.~(5): Continuum flux density at 3.3mm. We notice that only GN20 is
detected in the 3.3mm continuum. For GN20.2a and GN20.2b we list 2-sigma upper limits. 
The formal 3.3mm continuum measurement are: $0.11\pm0.10$ and $0.08\pm0.09$, respectively.
Col.~(6): VLA 1.4~GHz
flux density from P06$^{\dagger}$ and our own measurements. 
Col.~(7): {\it Spitzer}+MIPS 24~$\mu$m flux density. 
Col.~(8) and (9): {\it Spitzer}+IRAC flux densities. All detections have high significance, errors are dominated
by systematics effects at the 10\% level.
Col.~(10): Velocity integrated CO[4-3] flux. 
Col.~(11): Width of CO[4-3] line, estimated through a single Gaussian fitting.
}
\label{tab:GN20}
\end{deluxetable*}

\subsubsection{Photometric redshift from the infrared and radio emission}
\label{sec:chi2IR}

An independent estimate of the redshift of GN20 can be derived from 
its FIR emission. The ratio between mid-IR, FIR, mm and
radio flux densities is redshift dependent. This is due to the fact
that the radio luminosity is observed locally to be proportional to
the bolometric IR luminosity (Condon et al. 1992; Yun et al. 2001),
which seems to hold at high redshift as well (Elbaz et al. 2002; 
Garrett 2002; Appleton
et al. 2004). Originally this idea was exploited by Carilli \& Yun
(1999; 2000) to estimate photometric redshifts of distant SMGs,
also attempted by several other studies
(Hughes et al. 2002; Wiklind 2003; Aretxaga et al. 2003; 2005; 2007;
Clements et al. 2008).  In addition, one can exploit the fact that the
mid to FIR emission has a generally well defined peak at
$50-200\mu$m, affecting directly the ratio of mid-IR to submm and mm
emission.

For this purpose, we use the available photometry at 24$\mu$m, 850$\mu$m,
3mm and 20cm. Some of the {\it Spitzer}+MIPS 24$\mu$m and VLA 20cm
measurements are updated from P06 using the
most up-to-date GOODS datasets. We measure a 24$\mu$m flux density of
$65.5\pm3.5\ \mu$Jy and a 20cm flux density of $75.8\pm7.9\ \mu$Jy. We
use a 850$\mu$m flux density of $20.3\pm2.3$~mJy from P06 and a 3.3mm
flux density of $0.33\pm0.07$~mJy (Section~\ref{sec:pdbi}).

The observed flux densities have been compared, as a function of redshift, to the predictions of a suite of 105
template SEDs that were built following the luminosity correlations observed for local
galaxies as described in  Chary \& Elbaz (2001). 
The template dust
temperatures get warmer with increasing luminosity from a minimum template luminosity of 
${\rm Log}\ L_{\rm IR}/L_\odot=8.41$ up to a maximum
luminosity of ${\rm Log}\ L_{\rm IR}/L_\odot=13.55$.  
Radio continuum emission is added to each template following the radio-IR correlation (Yun et al. 2001).
For each template of a given luminosity we compute the expected flux densities in the available
bands (24$\mu$m, 850$\mu$m, 3mm and 20cm) as a function of redshift
and perform a $\chi^2$ minimization over the template total IR
luminosity as a function of redshift, without allowing template normalizations to vary. 
In order to avoid biasing the fitting
to a particular band, we increased the formal uncertainty in the
24$\mu$m flux density to obtain a $S/N$ ratio of 9 
(as opposed to the higher measured value of $\sim20$), comparable to the radio and submm bands.

Fig.\ref{fig:chi2F} shows the results. The best fitting redshift is
$z_{\rm phot}=3.7$ and at the 99\% confidence level is $3.3<z<4.3$.
Therefore, we conclude that the radio-IR SED also independently
supports the $z=4.055$ redshift determination. We
notice that this analysis allows us to reject with high confidence any
redshift identification with $z<3$ or $z>5$.

Fig.\ref{fig:chi2F} (right panel) also shows the comparison of the
observed GN20 SED from UV to radio to the best fitting (and brightest)
template in the library of Chary \& Elbaz (2001). The SED is
remarkably well reproduced in the mid- to FIR and radio, without
any rescaling or {\em ad-hoc} normalizations. The UV-optical part
of the template SED also happens to match the shape of the observed
data quite well, although there is a factor of $\sim2$ offset over
the IRAC bands that implies a different $L({\rm FIR})/L({\rm
optical})$ ratio, suggesting a higher specific SFR (SSFR).  A similarly
good agreement, including the optical/UV part, was seen by Daddi
et al. (2005b) when comparing the average SED of $z=2$ BzK selected
ULIRGs to Chary \& Elbaz (2001) templates. The optical/UV emission
was simply added to the models using a synthetic stellar population
model matched to the Arp220 SED and scaled to reproduce the local
broad correlation between bolometric luminosity and $B$-band rest
frame luminosity.

\subsubsection{Keck spectroscopy}
\label{sec:kspec}

A solid identification of the CO line, and thus a
determination of GN20's redshift, is finally provided by
deep optical spectroscopy.  The counterpart of GN20 was observed with
the Deep Imaging Multi-Object Spectrograph (DEIMOS; Faber et al. 2003)
on the Keck~II telescope several times between 2004 and 2007, though
only the final observations on UT 2007 April 14$-$15 were in good
conditions.  The total integration time was 2.5~hr, and a 1\farcs2
wide slit was oriented along the major axis of the galaxy (PA $=
13.8^\circ$).  The 600ZD grating ($\lambda_{\rm blaze} = 7500$ \AA)
was used, providing a resolution $R = 2000$.  The GG455 order-blocking
filter was installed.  We processed the data using a slightly modified
version of the DEEP2 DEIMOS pipeline\footnote{See {\tt
http://astron.berkeley.edu/$\sim$cooper/deep/spec2d/}.}.

The data show no emission lines while the continuum has a low $S/N$
per pixel, as expected given the faintness of the source ($i\sim25$)
and its spatial extent of about 1.5$''$, implying a low surface
brightness.  The emission is nevertheless clearly seen after smoothing
the spectrum.  We extracted the galaxy spectrum with {\em apall}
under IRAF using a $1.2''$ wide aperture.  We measured the r.m.s.
per pixel as a function of wavelength based on the sky spectrum.
This information was used to properly weight pixels when binning
by wavelength (e.g., to down weight noisy pixels affected by OH sky
lines) and to accurately estimate the error of the binned spectrum.
Given that for $z=4.055$ the Ly$\alpha$ forest break would fall at
6145\AA\, we included two spectral bins bracketing this wavelength
to verify if a break is indeed present as expected. Fig.~\ref{fig:chi2P}
(right panel) shows the results. Although the Keck spectrum only
extends down to 5900\AA, and therefore only one binning point is
available with decent $S/N$ below the expected Ly$\alpha$ forest
break, we find that the spectrum strongly supports the $z=4.055$
redshift and in particular it reveals the presence of the expected
Ly$\alpha$ forest break.  The flux density measurement blueward of
the expected break corresponds to a Ly$\alpha$ forest decrement of
$D_A = 0.55 \pm 0.10$ (derived using the stacked bins closer to the break), 
fully consistent to what is expected for the mean
transmission of the intergalactic medium at $z=4.05$ (Songaila et
al. 2004).  When running a photometric redshift analysis on the
binned Keck+DEIMOS data (Fig.~\ref{fig:chi2P}, left panel) we see
that of the possible CO line transition identifications, the $z=4.055$
redshift is unambiguously chosen with very high confidence.

\begin{figure*}
\centering
\includegraphics[width=16.0cm,angle=0]{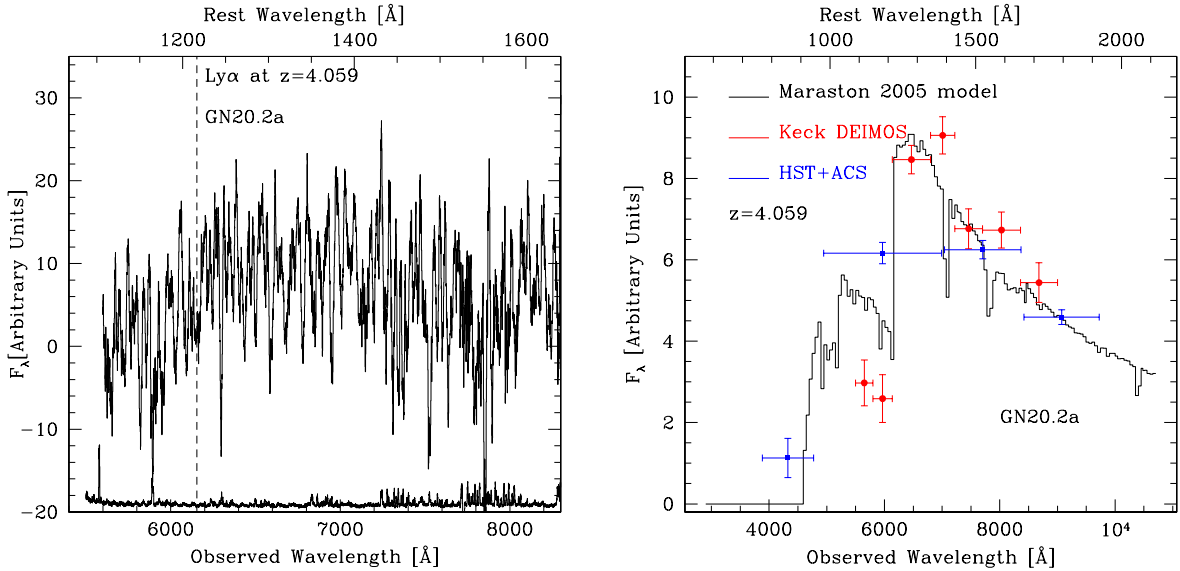}   
\caption{Keck spectroscopy results for the GN20.2a optical counterpart.
The left panel shows the Keck spectrum, smoothed with a 17\AA\ boxcar
(27 pixels) in order to increase the $S/N$. The Ly$\alpha$ forest break
is well detected, yielding a redshift of $z=4.059\pm0.007$ (1$\sigma$,
determined from spectral fitting with a Maraston [2005; M05] template). This is consistent 
with $z=4.051\pm0.003$ derived from the CO emission.
The relative
sky noise versus wavelength is shown at the bottom of the plot. The
right panel  is as in Fig.~\ref{fig:chi2P}, but for the GN20.2a galaxy,
showing the binned Keck spectra and the {\it HST}
photometry. 
The best fitting M05 model is shown, redshifted to
$z=4.059$. 
}
\label{fig:keck_GN20.2}
\end{figure*}

\subsection{GN20.2a}

A nearby ($24''$ separation; 169~kpc proper at $z=4.05$)
SMG companion to GN20 is GN20.2, first discovered by Chapman et al.
(2001a)\footnote{Chapman et al. (2001a) used SCUBA to observe (and
successfully detect) objects from a sample of optically faint radio
sources with $I>$25~mag. GN20.2, alias VLA~J123708$+$622201, was
part of this survey.}. GN20.2 has an 850$\mu$m flux density of 
9.9~mJy as reported by P06, that also identifies the most likely
optical counterpart with a relatively bright radio source ($S(20{\rm
cm})=180\ \mu$Jy). This galaxy is also a B-drop (P06; Fig.~\ref{fig:ACS})
with a faint 24$\mu$m flux density of $30.2\mu$Jy.

An analysis similar to the one performed in Sect.\ref{sec:GN20}
results in a photometric redshift of $z_{\rm phot, opt} = 3.93$ and
a 90\% confidence range within $3.55<z<4.05$. Similarly, the analysis
of the IR SED suggests a redshift of $3.4<z<4.2$ at the 99\%
confidence. This confirms that the overall properties of this source
are compatible with being at the same redshift of GN20. This is
further demonstrated by deep Keck+DEIMOS spectroscopy (2.5~hours)
of this galaxy that was obtained in April 2005 during cirrus
conditions, and reduced in the same way as for the GN20 observations
(see Sect.~\ref{sec:kspec}).  Fig.~\ref{fig:keck_GN20.2} shows that
a strong spectral break is detected at about 6150\AA, consistent
with the location expected for the Ly$\alpha$ forest break at a
redshift of 4.055. The observed break corresponds to a Ly$\alpha$
forest decrement $D_A = 0.67 \pm 0.05$, larger than the average
expectations but still within the observed scatter in the intergalactic
medium transmission at $z=4.05$ (Songaila et al. 2004).  We constrain
the redshift of this source to be $z=4.059\pm0.007$ ($1\sigma$),
based on fitting the Keck spectroscopy with the Maraston (2005)
 model shown in Fig.~\ref{fig:keck_GN20.2}, demonstrating
that this source is at a similar redshift of its brighter
companion GN20. We conclude that also in this case the detected CO
emission is CO[4-3] at $z=4.051\pm0.003$, in good agreement with
the Keck spectroscopy.

\subsection{GN20.2b: redshift and contribution to the submm flux}
\label{sec:sub_multiple}

The GN20.2b radio source is 3.4$''$ from the
GN20.2 SCUBA position of P06, closer than GN20.2a which is $6.0''$ (although,
we caution, that the SCUBA position of GN20.2 might be less accurate
than for the average SMG, due to the brighter and nearby companion,
GN20). The radio position is close (only 0.5$''$ to the north) to
a very faint galaxy ($z=27.34$) detected in the {\it
HST}+ACS imaging (Fig.\ref{fig:ACS}).
It is possible that this faint radio galaxy could be at the same $z\sim4.05$ redshift of GN20 and GN20.2a,
and it could be contributing some of the submm flux detected at the position of GN20.2. We will discuss these
two issues in turn in the following.

Although undetected in the F435W band, the faint ACS counterpart to GN20.2b is not formally
classified as a $B$-band dropout due to the upper limit in the F435W
not being stringent enough, but its blue $(i-z)$ and red $(v-i)$
colors support the hypothesis that also this object is at $z\sim4$.
The galaxy is also fairly bright and red at IRAC wavelengths (SED
peak at 8.0$\mu$m with 20.93 mag) and is selected as a $z>3.5$
massive galaxy candidate in the work of Mancini et al. (2008).
Inspecting the highest resolution IRAC imaging at 3.6$\mu$m and 4.5$\mu$m it appears that 
blending with a neighbouring source is somewhat affecting the {\em Spitzer} IRAC measurements and 
biasing the flux densities high (likely by at most a factor of 1.5--2).
We used the radio-IR
photometric redshift technique to constrain its redshift, in the
assumption that the emission is dominated by star formation and not
an AGN.  We find that by itself, the large radio to 24$\mu$m flux
density ratio for this source strongly constrains its redshift to
$z>3.2$ (99\% confidence level), independently of how much it
contributes to the 9.9mJy flux density at 850$\mu$m.  This is
consistent with this object being an intrinsically very luminous,
highly star forming galaxy at high redshift (although, in principle,
the high ratio could be also due to radio-loud AGN emission from
lower redshift).  
As discussed in Sect.~\ref{sec:20p2b}, there is possible evidence for CO
emission from this source, where a positive signal is found at the
$>4\sigma$ level.  Although the situation is not secure as for
the other two sources, we conclude that there is some supporting evidence
for a similar $z\sim4.05$ redshift also for GN20.2b. Taking the CO
signal at face value, we would assign a redshift of $z=4.052\pm0.006$
to GN20.2b, where the error corresponds to half the frequency range
spanned by the tentative positive emission, if we identify that as CO[4-3]. 
This is within $\Delta z=0.003$ of the redshift of GN20 and GN20.2a, 
while our data is sensitive in principle to emission from a $\Delta z=0.05$, 17 times larger
redshift range of $4.02<z<4.07$.
We conclude that it is at least plausible that GN20.2b
could be lying at a similar redshift of GN20.2a. In any case, this appear to be a fairly
high redshift ($z>3.5$) galaxy.

Could GN20.2b be contributing some major fraction of the
submm flux of GN20.2 ? 
Finding multi-component systems is not a new situation for SMG
identification --- e.g., SMMJ094303 (Tacconi et al. 2006) and GN19
(HDF242; Tacconi et al. 2008) are two cases where two-component
systems, both radio detected, have been confirmed through CO
observations.  Dannerbauer et al. (2004), P06, Ivison et al. (2007),
and Younger et al. (2008a) report other cases of submm/mm galaxies
with two radio counterparts. If this is also happening in 
the present case of GN20.2 is not easy to assess quantitatively.
We present below
circumstantial evidences in favor or against this hypothesis.
%The IRAC flux densities of GN20.2b
%are comparable to those of GN20 and $2-3$ times brighter than those
%of GN20.2a, and its ACS to IRAC colors are correspondingly redder,
%suggesting that GN20.2b is more massive than GN20.2a if truly at $z>3.5$.
%Accounting for blending in the IRAC photometry of GN20.2b is likely to 
%reduce the difference in the stellar mass, still without removing it.

The main supporting evidence is that
its VLA 1.4~GHz radio continuum
flux density is $42\pm10$\% that of GN20, which is similar to the ratio between the submm fluxes of
the same two galaxies,
suggesting that GN20.2b might be contributing some substantial flux, unless
the radio emission is affected by an AGN. We notice though that
GN20.2b is 5.5 times fainter in the VLA 1.4~GHz radio continuum than GN20.2a, but this
is likely due to the fact that the radio emission of GN20.2a (even more than twice larger than
that of GN20) is powered by an AGN,  as discussed in the following section. 

Evidences disfavoring instead a major submm flux contribution are the following.
Comparing the 24$\mu$m flux density measurements,
we find that GN20.2b (only tentatively detected at 24$\mu$m) is at least 2--3 times fainter at 24$\mu$m than GN20.2a
and at least 5.5 times fainter than GN20. Accounting also for blending, the  $S(24\mu {\rm m})/S(4.5\mu {\rm m})$ 
flux density ratios are at most a factor of 2 for GN20.2b and about factors of 6--7 for GN20 and GN20.2a
(Tab.~\ref{tab:GN20}). This is evidence for lower specific SFR in GN20.2b, suggesting
that this galaxy has already passed its major peak of activity. 
Similarly, the ratio of the CO fluxes detected for GN20.2 and GN20.2a is roughly a factor of 2, similar to 
the ratio of submm fluxes between the two galaxies (Tab.~\ref{tab:GN20}).  Even if the tentative CO emission of 
GN20.2b is real, this would still be at most half of the CO emission of GN20.2a (but, of course, there could be 
place for stronger CO emission if the galaxy lies outside of $4.02<z<4.07$). 
Given that the CO and bolometric luminosities of SMGs are known to correlate (Greve et al. 2005; Solomon \&
van den Bout 2005; see also Section~\ref{sec:CO}), this could suggest that GN20.2a
does account for most of the submm emission of GN20 from P06, and that GN20.2b could contribute at most some 30\%
of it but likely not much more. A final argument is that GN20.2 was originally discovered by Chapman et al (2001a)
in {\em photometry} mode with SCUBA at JCMT, 
pointing at the position of the GN20.2a radio galaxy. This resulted in a submm flux
density of $S(850\mu {\rm m})=10.2\pm2.7$ that is fully consistent with the flux in the P06 map. GN20.2b is 6.8$''$
away from GN20.2a and any emission from its position would have been close to the edge of the telescope primary beam,
resulting attenuated by a factor of 2. Nevertheless, the Chapman et al (2001a) flux is actually slightly brighter
than the flux in P06, although consistent within the errors.
Accounting  for the errors in the P06 and Chapman et al. (2001a) $850\mu$m
flux density measurements, this suggests that GN20.2a does contribute at least half of the flux of GN20.2,
and possibly most of it.

\begin{figure}
\centering
\includegraphics[width=8.8cm,angle=0]{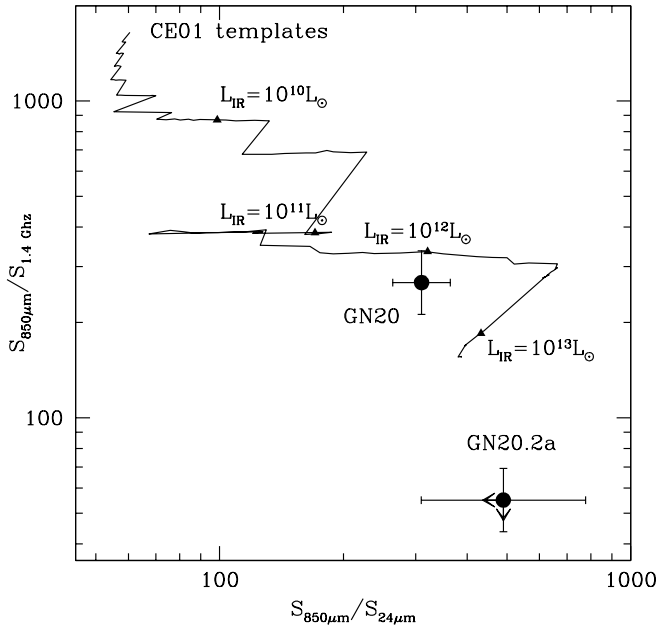}   
\caption{The $850 \mu$m to $24 \mu$m and 1.4~GHz flux density
ratios for GN20 and GN20.2a, filled circles. We assign to GN20.2a
the full $850 \mu$m flux density of GN20.2 although GN20.2b might 
be contributing some of the submm flux, as discussed in Sect.~\ref{sec:sub_multiple}.
In that case, both flux density ratios should be taken as upper limits.
The lines show the predicted ratios labeled as a
function of bolometric luminosity ($L_{\rm IR}$) for the CE01 library
of templates at a redshift of $z=4.055$.
}
\label{fig:colorS}
\end{figure}

%Given that the radio flux density of GN20.2b is roughly half that
%of GN20 and the 850$\mu$m flux density of GN20.2 is 9.9~mJy (P06)
%or roughly half that of GN20, it is plausible that also this source
%could be at the same $z\sim4.05$ redshift and contributing some
%submm emission. The marginal detection at $24\mu$m (measured flux
%density of 12$\mu$Jy, significant at the $\sim3\sigma$ level) is
%still consistent with this hypothesis.  
For completeness, we notice that an additional faint radio-detected galaxy is present just 1$''$
North outside the edge of the SCUBA beam. This galaxy is a hard X-ray source in the castalog
of Alexander et al. (2003), for which we derive a photometric redshift of $z_{phot}=1.6$. 
The radio and mid-IR emission of this object, if entirely due to star formation would correspond 
to about 100~M$_\odot$~yr$^{-1}$. This in turn would produce at most a submm flux of order of 1~mJy at 850$\mu$m.
Being at 8$''$ from the P06 position and at 14$''$ from the Chapman et al. (2001a) pointed observations, 
any potential submm emission from this galaxy would be negligible for the GN20.2 system 850$\mu$m photometry.

In conclusion, 
our results suggest a scenario in which two $z\sim4.05$
counterparts might be contributing to the submm emission of GN20.2, GN20.2a and GN20.2b, although GN20.2a is likely 
accounting for most of the submm emission.

\section{The physical properties of $z=4.05$ SMGs}
\label{sec:CO}

\subsection{Bolometric IR luminosities}

Following the analysis described in Sec.~\ref{sec:chi2IR}, the radio-IR SED of GN20 is best reproduced
with a CE01 library template corresponding to a luminosity 
of $L_{\rm IR}=3.5\times10^{13}L_\odot$, redshifted to $z=4.055$. However, the
CE01 library implies a correlation
between dust temperature and luminosity 
that is calibrated on observations in the local
Universe but that might not necessarily hold at higher redshifts. In order to explore the uncertainties 
in the determination of GN20's total IR luminosity due to possible SED temperature variations,
we used again the CE01 library, but allowing this time for a free normalization of the templates when comparing to the
observed fluxes, thus spanning the full range of dust temperatures in the
library. 
For this exercise, we used the measurements or upper limits at 24$\mu$m, 850$\mu$m, 3.3mm
and 20cm and included also the 1.1mm and 1.2mm measurements from Perera et al (2008) and Greve et al. (2008)
and the 70$\mu$m upper limit from Frayer et al. (2006).
Integrating over the best fitting template yields in this way $L_{\rm IR}=2.4\times10^{13}L_\odot$ and a formal
uncertainty from the fit at $1\sigma$ of less than 0.05~dex (the luminosity step in the CE01 library). 
The best fitting template has an intrinsic $L_{\rm IR}=1.5\times10^{12}L_\odot$ and is scaled up in luminosity by
a factor of 16. This would correspond to an SED temperature of about 57~K,
lower than those of CE01 templates of comparable luminosities $L_{\rm IR}\simgt10^{13}L_\odot$.
The same effect can be appreciated more directly in Fig.~\ref{fig:colorS} where 
we show the $850\mu$m to 24$\mu$m and radio flux density ratios.
The GN20 flux density ratios are reasonably in line with CE01
predictions, although the 850$\mu$m to radio flux density ratio is
80\% higher than expected for the $L_{\rm IR}=10^{13.5}L_\odot$
galaxy in the CE01 library. The effect is significant at the 2-$\sigma$
level only, but indeed its SED is closer to the $L_{\rm
IR}=10^{12}L_\odot$ expected ratio. 
SMGs have been generally found to be somewhat colder than CE01 models, as a result
of the selection at 850$\mu$m (P06), so this result is not too surprising.

As a third estimate of the IR luminosity of GN20 we might use directly 
the radio-IR correlation (the other two estimates also used this correlation to fit the radio, although this was 
weighted together with all the other measurements).
For its measured 1.4~GHz flux density of $75.8\mu$Jy, assuming a radio continuum
$(f_{\nu}\propto \nu^{-\alpha})$
with $\alpha=0.8$, 
we would derive $L_{\rm IR}=2.8\times10^{13}L_\odot$. 
It is not obvious to decide which of the 3 estimates is more reliable but fortunately the measurements are very close, 
all within a range of less than 0.2~dex.
In summary, we adopt the average value of $L_{\rm IR}=2.9\times10^{13}L_\odot$ for our best estimate and an uncertainty
of 0.2~dex.

The 850$\mu$m to VLA radio flux density
ratio for GN20.2a is about $5\times$ lower than that for GN20
(Fig.\ref{fig:colorS}), 
implying a strong radio excess
of a factor of $4-5$. 
GN20.2a has a radio
luminosity of $2\times10^{25}$~W~Hz$^{-1}$ at 1.4~GHz rest frame,
using again a radio continuum index $\alpha=0.8$.
This demonstrates that a powerful
radio-loud AGN is hidden in this $z=4.05$ source. Therefore, we cannot use the 1.4~GHz flux density to
estimate its total IR luminosity. We used the CE01 library templates, with and without allowing for a 
free normalization, to fit the 24$\mu$m, 70$\mu$m, 850$\mu$m and 3.3mm measurements or upper limits. 
Similarly to GN20, we find $L_{\rm IR}=2.2\times10^{13}L_\odot$  and $L_{\rm IR}=1.0\times10^{13}L_\odot$, the
latter being the 'free-normalization' case that also prefers a lower temperature template. These estimates 
use the full 850$\mu$m flux density measurements although some part of this might be due to GN20.2b. This 
also assumes that the AGN inside GN20.2a is not contributing substantially to the measured 24$\mu$m and 850$\mu$m
flux densities. Pope et al. 2008 showed that this is generally the case at least for $z\sim2$--3 SMGs.
Given the uncertainties, we adopt the average of these two estimates 
$L_{\rm IR}=1.6\times10^{13}L_\odot$ and an uncertainty of about 0.3~dex. 

This analysis implies that GN20 and GN20.2  are extremely luminous galaxies,
with $L_{\rm IR}>10^{13}L_\odot$. 
Besides being among the most distant known SMGs, these are also some of
the most luminous galaxies known so far.
We use a Kennicutt et al.  (1998) conversion 
of $SFR[{\rm M}_\odot$~yr$^{-1}] = L_{\rm IR}[L_\odot]/10^{10}$,    
expressed for the Chabrier 2003 IMF adopted in this paper.
The IR luminosities correspond to SFRs of
$>1000M_\odot$~yr$^{-1}$ 
if the IR emission is dominated
by star formation.  We note that neither the GN20 nor GN20.2a (or GN20.2b)
counterparts are detected in the X-rays in the catalog of Alexander et al. (2003), 
although the radio emission does suggest 
the presence of an AGN inside GN20.2a. There is no clear evidence, instead, for the
presence of an AGN inside GN20.
If a top-heavy IMF
is adopted, the implied SFR could also become significantly smaller.

\begin{figure*}
\centering
\includegraphics[width=16.0cm,angle=0]{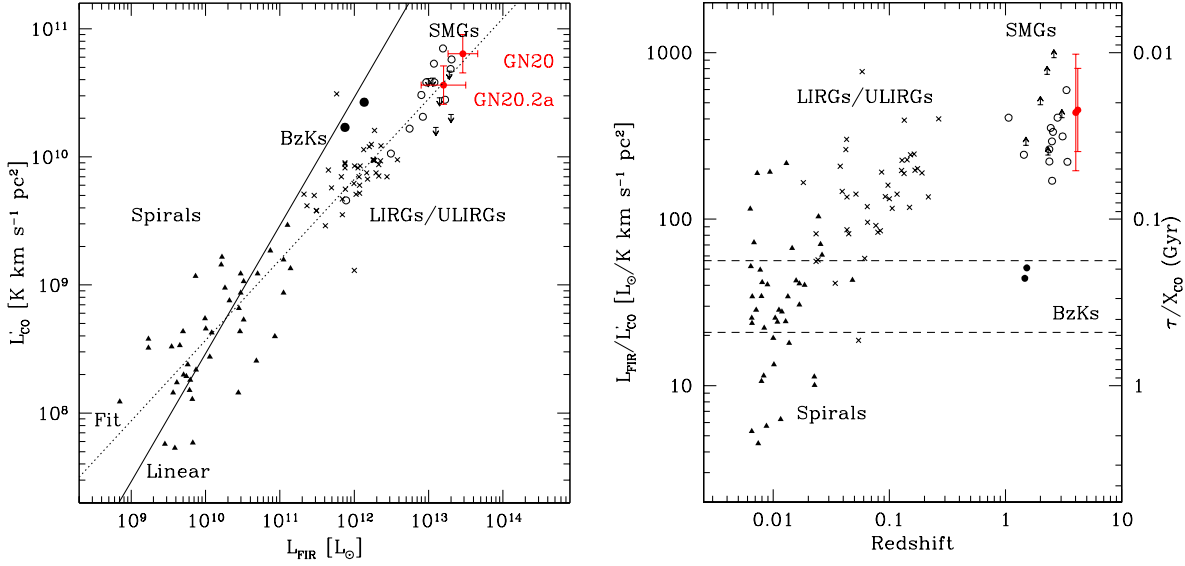}   
\caption{Comparison of FIR to CO luminosities for a variety of local
and distant galaxies (see Solomon \& van den Bout 2005; Greve et
al. 2005; figure adapted from Daddi et al. 2008) and the GN20 and
GN20.2a galaxies (points with error bars). The FIR estimates for GN20
and GN20.2a are described in the text, based on the radio-IR SEDs.  
The CO luminosities are estimated from the CO[4-3] transitions
for GN20 and GN20.2a, and for a variety of other transitions for
the other sources. We assume constant brightness temperature in the
various CO transitions for this comparison.  Notice that CO luminosities
(and therefore gas masses) could be underestimated if the observed CO transitions
are not thermalized.
}
\label{fig:CO_FIR}
\end{figure*}

\begin{deluxetable*}{lcccccc}%[h]
\tabletypesize{\scriptsize}
\tablecaption{Derived properties of GN20 and GN20.2a}
\tablewidth{0pt}
\tablehead{
\colhead{ID} &
\colhead{$L_{\rm IR}$}&
\colhead{$L'_{CO[4-3]}$}&
\colhead{$L_{1.4~GHz}$}&
\colhead{$M_{stars}$}&
\colhead{$M_{gas}$}&
\colhead{$M_{dyn}$}
\\
\colhead{}&
\colhead{$L_\odot$}&
\colhead{K~km~s$^{-1}$~pc$^2$}&
\colhead{W~Hz$^{-1}$}&
\colhead{$M_\odot$}&
\colhead{$M_\odot$}&
\colhead{sin~$i^{-2}$~$M_\odot$}
\\
\colhead{(1)}&
\colhead{(2)}&
\colhead{(3)}&
\colhead{(4)}&
\colhead{(5)}&
\colhead{(6)}&
\colhead{(7)}
}
\startdata
GN20 & $2.9\times10^{13}$ & $6.2\times10^{10}$ & $8.4\times10^{24}$ & $2.3\times10^{11}$ & $5.0\times10^{10}$ & $2.3\times10^{11}$\\
GN20.2a & $1.6\times10^{13}$ & $3.7\times10^{10}$ & $2\times10^{25}$ & $0.5\times10^{11}$ & $3.0\times10^{10}$ & --
%GN20.2b & $\sim 1.2\times10^{13}$ & $1.8\times10^{10}$ ? & $3.5\times10^{24}$ & $1.2\times10^{11}$ & $1.5\times10^{10}$ ? & --
\enddata
\tablecomments{
Col.~(1): Name of the object. 
Col.~(2): total IR luminosities for GN20 and GN20.2a
are derived by fitting the global IR SED using all available flux density
measurements from 24$\mu$ to 1.4~GHz with CE01 models. For GN20.2a, the 1.4~GHz measurement was not used, as 
it is affected likely by an AGN, and we attributed the whole 850$\mu$m emission to this source. 
Typical errors are estimated to be of order of 0.2~dex and 0.3~dex for GN20 and GN20.2a, respectively. 
Col.~(3): luminosity of the CO[4-3] transition. 
Col.~(4):  1.4~GHz rest frame luminosities, estimated
using a radio continuum index $\alpha=0.8$ $(f_{\nu}\propto \nu^{-\alpha})$, $L_{1.4 {\rm GHz}} = 4\pi D_L^2 S(1.4 {\rm
GHz}) (1+z)^{-0.2}$ ($D_L$ is the luminosity distance).
Col.~(5): stellar masses are estimated from SED fitting from HST+ACS to Spitzer+IRAC photometry, using constant 
star formation rate models from Maraston (2005) and allowing for dust reddening with a Calzetti et al. (2000) law.
Typical errors are found to be about 0.2~dex.
Col.~(6): molecular gas masses assume  that the CO[1-0] and CO[4-3] transitions correspond to
the same brightness temperature and a conversion factor $X_{\rm
CO}=0.8$~$M_\odot$~(K~km~s$^{-1}$~pc$^2$)$^{-1}$.
Col.~(7): the dynamical mass for GN20 is estimated following Solomon \& van den Bout (2005) using 
$M_{dyn}\times{\rm sin} i^{2}/M_\odot=233.5\times v_{\rm FWHM}^2\times r_e$ (with velocity expressed
in km~s$^{-1}$ and half light radius in pc).\\
%All quantities reported for GN20.2b assume a redshift of $z=4.05$ that, we recall, is not yet solidly confirmed.
%A question mark close to CO derived quantities reminds that the CO detection of this galaxy is only tentative.
}
\label{tab:GN20_est}
\end{deluxetable*}

\subsection{CO luminosities and molecular gas masses}

The observed CO[4-3] fluxes convert to luminosities of $L'_{\rm CO[4-3]}=6.2\times10^{10}L_\odot$
and $L'_{\rm CO[4-3]}=3.7\times10^{10}L_\odot$, for GN20 and GN20.2, respectively. 
Fig.\ref{fig:CO_FIR} shows that these galaxies appear
to lie close to the (non-linear) correlation between $L'_{\rm CO}$ and $L_{\rm IR}$ traced
by local ULIRGs and distant SMGs. 
In the
assumption that the CO[1-0] and CO[4-3] transitions correspond to
the same brightness temperature and assuming $X_{\rm
CO}=0.8$~$M_\odot$~(K~km~s$^{-1}$~pc$^2$)$^{-1}$ (Downes \& Solomon
1998; Solomon \& van den Bout 2005), the CO luminosities convert to total molecular
gas masses of $5\times10^{10}M_\odot$ and $3\times10^{10}M_\odot$ for GN20 and GN20.2a, respectively. 
These might be regarded as lower limits as, in the case that the CO[4-3] transitions observed were not thermalized,
the total CO luminosities and gas masses would be higher. VLA observations of CO[1-0] will be able to address this point.

At the observed SFR levels, the gas reservoirs would be exhausted in roughly 
20--30~Myr (Fig.\ref{fig:CO_FIR}).
Similar timescales have been derived for $z\sim2$--3 SMGs by Greve et al. (2005).

\subsection{Dynamical and stellar mass estimates}

The CO[4-3] emission from GN20 has a FWHM of 710$\pm120$~km s$^{-1}$,
matching well to the typical value of CO-detected SMGs at lower
redshifts (Greve et al. 2005). From the higher resolution B~configuration
PdBI data alone we find that the CO line is marginally resolved spatially,
with a FWHM of $0.42''\pm 0.18''$ from Gaussian fitting of the {\em uv} visibilities performed within GILDAS. 
The possible
extension of the source is supported by the fact that for point source 
extractions, the flux measured with the D-configuration data is higher than the one
from the B-configuration data, a $2\sigma$ effect suggesting that the CO emission
is starting to be resolved at the 1.3$''$ resolution of our B-configuration data.
This size matches well to the typical
CO sizes of $r_e\sim2$~kpc for SMGs (Tacconi et al. 2006; 2008),
and is roughly consistent with its UV extension as well (Fig.~\ref{fig:ACS}).
Recently, Younger et al. (2008b) measured a size of $0.6''\pm0.15''$
for GN20 (Gaussian fitting), from 
the far-IR dust continuum emission at 890$\mu$m with the Sub Millimeter Array,
which is fully consistent with our CO size estimate.
The velocity and size correspond to a total dynamical mass of $\sim
2.3\times10^{11} {\rm sin} i^{-2}M_\odot$ including gas, stars and dark matter.

\begin{figure*} \centering
\includegraphics[width=16.0cm,angle=0]{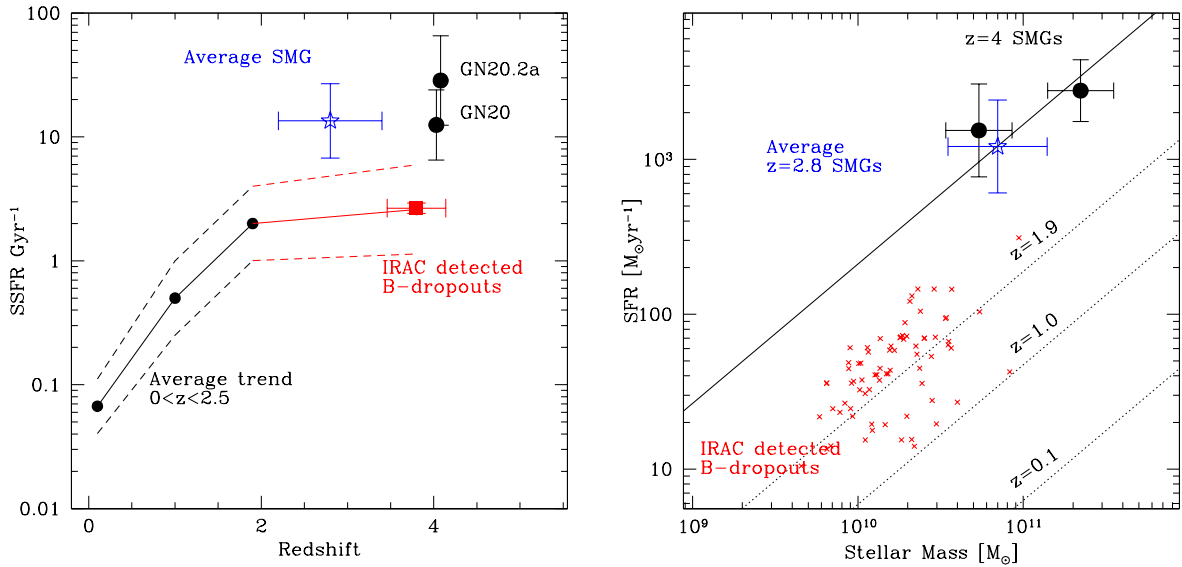}
\caption{Specific star formation rate (SSFR; SFR to stellar mass ratio) as a function of
redshift is shown in the left panel. The right panel shows the
stellar mass versus star formation rate.  Large solid circles are
GN20 and GN20.2a.  Their stellar masses were derived through SED fitting, and typical 
errors are found to be of the order of 0.2~dex. 
The filled square (left panel) and small crosses (right panel) refer to measurements
derived for the IRAC detected, massive $B$-band dropout Lyman break
galaxies. Stellar masses and SFRs of the B-band dropouts were obtained through fitting of the UV to IRAC SEDs. 
Individual measurements are shown in the right panel, while the left panel shows B-dropout's 
average quantities
(with error bars showing the error in the mean of the SSFR of the ensemble and the expected range
in redshift).
Stacking at 850$\mu$m and 1.4~GHz yield upper limits on the average SFR of order of $55 M_\odot$~yr$^{-1}$
for the B-band dropouts,
consistent with the average UV measurements. The star refers to the
average of SMGs at $z\sim2$--3 from Tacconi et al. (2006); error bars refer to the spread of properties in the sample.
Filled black
circles in the left panel show the average SSFR versus redshift at
a stellar mass of $5\times10^{10}M_\odot$ from Daddi et al. (2007a) and
Elbaz et al. (2007). The dashed lines show the measured 1-$\sigma$ range (about 0.2~dex
at $z=0.1$ and about 0.3~dex at $z=1$ and 1.9 and 0.36~dex at $z=3.8$).
We recall that stellar masses and SFRs in this figure are shown for the Chabrier (2003) IMF.
}
\label{fig:Mass_SFR} 
\end{figure*}

Given the molecular gas mass estimates,
this corresponds to a
molecular gas fraction of about 20\%, similar to what is typically
found for lower redshift SMGs (Greve et al. 2005).  The stellar
mass derived from SED fitting of the ACS to IRAC photometry is
$\sim2.3\times10^{11}M_\odot$, expressed for a Chabrier (2003) IMF
(see, e.g., Maraston et al. 2006 for more details on the range of models
and assumptions implied).
The combined molecular gas and stellar masses, only roughly half
of which presumably fall within $r_e$, add up to roughly 70-80\%
of the estimated dynamical mass, close enough within the large
uncertainties of such estimates.  
This implies that the stellar
mass is not highly overestimated for the choice of a Chabrier (2003) IMF
and the case of a top heavy IMF is disfavored by the data. 

For GN20.2a, we infer with a similar method a stellar mass of $\sim0.5\times10^{11}M_\odot$.

\subsection{On the possibility of lensing}

The intrinsic SFRs and luminosities of these sources might be lower  if
there is  amplification by gravitational lensing  (see Paciga \& Scott 2008). 
Lensing by foreground structure could be an alternative way to interpret the fact that colder SEDs, 
typical of lower luminosity galaxies in the
local Universe, seem to reproduce better the colors of the GN20 and GN20.2 galaxies. In that case, also
the luminosity would match well to those of colder galaxies, without the need to advocate 
evolution in the temperature-luminosity relation of starburst galaxies.
Some evidence against lensing is however provided by the FIR to CO properties
of GN20 and GN20.2 (Fig.~\ref{fig:CO_FIR}). 
If lensed by a factor of 10 or
more, the de-lensed properties would place them in a region of the
diagram corresponding to a very high star formation efficiency, not
commonly observed for $L_{\rm IR}\sim10^{12}L_\odot$ sources at high
and low redshifts.
Following the analysis discussed
in Tacconi et al. (2008), the agreement between the dynamical mass estimate 
and its stellar and gas mass estimates
also disfavors the possibility that GN20 is lensed by a very large factor,
given that emission line FWHMs should be independent on magnification. 
Qualitatively,
the same picture also holds for GN20.2a, for which the data suggest
a CO[4-3] line FWHM$\sim1000$~km s$^{-1}$ (albeit with a large
error), not uncommon amongst SMGs (Greve et al. 2005). 

\subsection{The relation between stellar mass and SFR at $z\sim4$}

A tight correlation exists between the SFR and stellar mass of star
forming galaxies from $z=0$ to $z\sim2$, after excluding the locus
of quiescent/passive galaxies (Elbaz et al. 2007; Daddi et al.
2007a; Noeske et al. 2007). The implication of this correlation
remains subject to debate but suggests that, {\it on average}, the
processes regulating the star formation activity of a galaxy are
the same over a large range of stellar masses. Outliers do exist
at both extremes in this correlation. On the low SFR side, the cloud
of red-dead galaxies have SFRs lower than expected by several orders
of magnitudes.  On the high SFR side, ULIRGs and $z\sim2$--3 SMGs
exhibit approximately one order of magnitude larger SFRs for their
stellar mass than the average galaxy of equivalent stellar mass
(Daddi et al. 2007a; Elbaz et al. 2007). 
Here, following Daddi et al 2007a, we use the Tacconi et al. 2008 estimate 
of SSFR for $z=2$--3 SMGs that are based on dynamical masses rather than stellar masses which should be a
more robust derivation of the total galaxy masses, and we converted the average $L_{\rm IR}=10^{13.1}L_\odot$ of the
sample into $SFR\sim1200$~M$_\odot$~yr$^{-1}$. Takagi et al. (2008) confirm higher SSFRs in SMGs,
respect to normal galaxies of the same masses, using stellar mass estimates, although finding SSFRs
a factor of 3 higher on average and a few possibly overlapping galaxies (see also Dannerbauer et al. 2006
for overlapping objects between the BzK and SMG samples).
Their SFRs are based on the radio and assume a radio-IR correlation with a different normalization
respect to the local one by a factor of about 2. 

On the low side,
the red-dead galaxies appear to fall short of SFR because of a lack
of molecular gas fuel, possibly due to negative feedback, the excess
SFR of $z\sim2$--3 SMGs remains to be understood. A possible explanation
is external triggering, e.g. by major mergers (Tacconi et al. 2006;
2008) as it is seen for local ULIRGs.

When compared to their closer $z\sim2$--3 siblings, $z\sim4$ SMGs
appear to be forming stars with an equivalently high SSFR.  However, the
stellar mass-SFR relation shows a continuous increase of SSFR with
increasing redshift, which is expected due to the higher gas mass
fractions at higher redshifts.  The question therefore arises as
to whether an SMG could represent the typical star forming galaxy
at $z\sim4$, contrary to their $z\sim2$ siblings 
which are atypical
compared to the average galaxy at their cosmic epoch, having much larger 
SSFR, see Fig.~\ref{fig:Mass_SFR}. 

In order to address this question, one needs to identify the locus
of average star forming galaxies at $z\sim4$ and then compare it
to the locus of $z\sim4$ SMGs.  We consider the sub-sample of
$B$-band dropout Lyman break galaxies in GOODS-N with IRAC magnitudes
at 5.8 and 8.0$\mu$m brighter than 24. The requirement of an IRAC
detection is necessary to reliably estimate the stellar masses. The
use of $B$-band dropouts ensures that we are concentrating on star
forming galaxies, with a negligible contribution from passive
galaxies. Within this limit, about 10\% of the $B$-band dropouts
to $z_{\rm AB}<27$ are selected, or 77 sources.
We fitted the ACS to IRAC SEDs of these B-band dropouts using Maraston (2005) models
with constant star formation, 
allowing the redshift to vary freely within $3.1<z<4.4$ (the 95\% confidence region 
expected for the redshift distribution of B-band dropouts) and allowing also for dust
reddening following a Calzetti (2000) law. In this way, we estimated 
stellar masses and
(basically UV-driven)
SFRs for each of the B-band dropout Lyman break galaxy. Typical errors
from the fit amount to factors of about 2 for both quantities.
From this we derive
an average stellar mass of $1.5\times10^{10}M_\odot$ and an average SFR of 40$M_\odot$~yr$^{-1}$. 
We performed
stacking at 850$\mu$m using the SCUBA super-map of Pope et al.
(2005) and a technique identical to what was done in Daddi et al.
(2005b; 2007a). We do not detect the galaxies, deriving a $3\sigma$
flux density limit of 0.54mJy, corresponding to 
57~$M_\odot$~yr$^{-1}$. From radio stacking, again
performed with a similar technique to Daddi et al. (2005b; 2007a),
we get a tentative detection with a peak flux of $1.7\pm0.6\mu$Jy.
The $3\sigma$ radio upper limit corresponds to a 
SFR of 54$M_\odot$~yr$^{-1}$,
consistent with the limit from submm stacking. Overall, the stacking limits appear to be consistent 
with the SFRs derived from SED fitting in these galaxies. This is similar to what found at $z\sim2$
where analogous estimates based on the UV luminosity of galaxies appear to agree well on average
with mid- and far-IR estimates (Daddi et al. 2007a).

The locus of typical, massive $z\sim4$ B-band dropouts does not
support a continuously increasing SSFR with redshift, suggesting
instead a plateau of the SSFR for $z \simgt 2$.  
The average SFR of 40~$M_\odot$~yr$^{-1}$  
places these 'typical' $z\sim4$ galaxies fairly close to the $z\sim2$
correlation.

The $B$-band dropout Lyman break galaxies exhibit an average stellar
mass lower than the $z\sim4$ SMGs.  This implies that not only do
$z\sim4$ SMGs present similar SSFRs to their $z\sim2$ siblings, but
they are also found to be outliers with respect to the average
star-forming galaxy at their epoch. Note that the similarity of
$z=2$--3 and $z=4$ SMGs is also supported by their equivalently large
FIR to CO ratios, corresponding to high star formation efficiencies
and rapid gas consumption timescales.

We hypothesize that this favors a triggering mechanism for the
activity of the $z\sim4$ SMGs, such as major mergers.  Indeed,
GN20.2a has a very nearby ($0.5''$ or 3.5 proper kpc at $z=4.05$)
$B$-band dropout companion (Fig.~\ref{fig:ACS}). GN20 has a clumpy
morphology that could be reminiscent of a merger (although, see
Bournaud et al. 2008). The 0.5$''$ offset between the optical light
peak and submm/radio and CO might also point to a complex geometrical
situation often found among mergers (e.g., the Antennae; Wang et al. 2004).  
The high gas densities of a merging event could also explain
why, contrary to what is found for typical massive galaxies at
$1.4<z<2.5$ (Daddi et al. 2007a; Dannerbauer et al.  2006), the
emitting region is completely opaque in the UV, similar to $z\sim2$--3
SMGs.  Indeed, we find that the radio-IR based SFR estimates of the $z\sim4$ SMGs are
a factor of $>10$ times larger than the UV-based ones.

\begin{figure*} \centering
\includegraphics[width=16.0cm,angle=0]{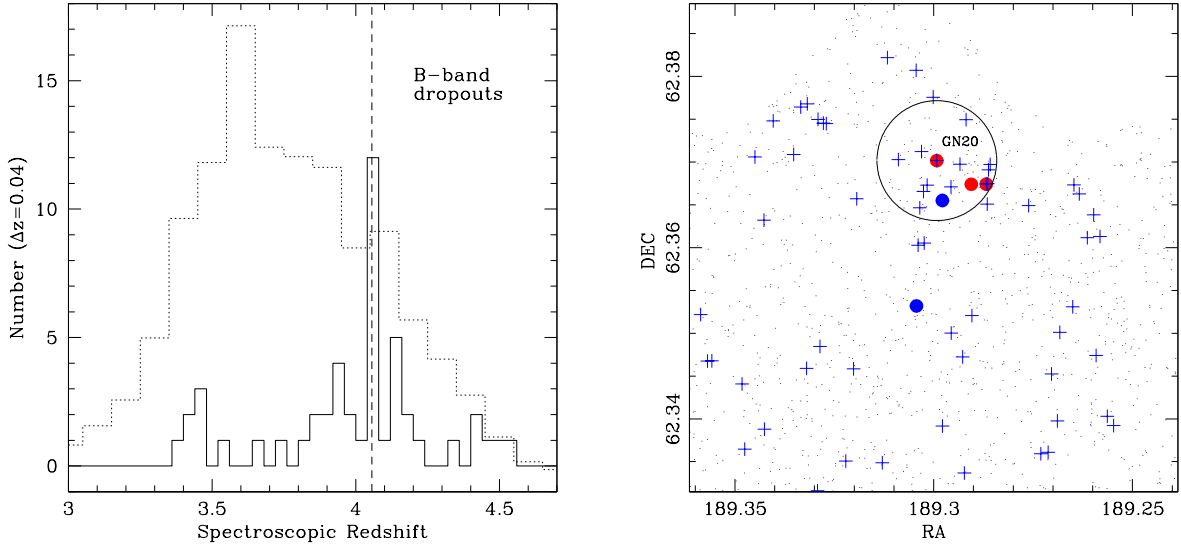} 
\caption{Left panel: spectroscopic redshift distribution of GOODS-N
$B$-band dropout Lyman break galaxies galaxies, in bins of 0.04
(solid histogram). A strong spike is seen, coincident with the
redshift of GN20 and GN20.2a (vertical dashed line). The dotted
histogram shows the expected redshift distribution of $B$-band
dropouts, based on the simulations of Giavalisco et al. (2004b) and
Lee et al. (2006).  Right panel: spatial distribution in the northern
corner of the GOODS-N field (about $3.5'\times3.5'$) of $z_{\rm AB}<27$
galaxies (small points), and $B$-band dropouts (crosses). The filled
symbols are GN20,  GN20.2a and GN20.2b. The large circle shows a 25$''$
(180~kpc) radius area centered on GN20.  The blue filled symbols
are two spectroscopically confirmed galaxies at $z=4.058$ (closer;
this is BD29079, also shown in Fig.~\ref{fig:ACS}) and $z=4.047$ (further).
} 
\label{fig:zhist} 
\end{figure*}

\section{A $z=4.05$ proto cluster of galaxies in GOODS-N centered around GN20 and GN20.2}
\label{sec:cluster}

Down to the limits of the GOODS-N observations of P06, the surface
density of SMGs is $\approx1000$~deg$^{-2}$, and thus the fact that
GN20 and GN20.2 are only $24''$ from each other, with a projected
spatial distance of 169 physical kpc,
and both are at $z\sim 4.05$, implies that these two SMGS are spatially
clustered (see also Blain et al. 2004).  This suggests that these
SMGs might also be part of some type of larger, high redshift
structure in GOODS-N.

To investigate further this possibility, we have analyzed the
distribution of spectroscopically confirmed Lyman-break galaxies
at $z\sim 4$ ($B$-band dropouts) in GOODS-N (Fig.~\ref{fig:zhist};
Stern et al., in preparation).  We compare the observed redshift
distribution to the expected distribution function derived from
simulations (Giavalisco et al. 2004b; Lee et al.  2006). There are
12 redshifts in the $\Delta z=0.04$ bin that includes GN20 ($z=4.055$);
these galaxies define a clear spike in the redshift distribution.
There currently are 58 high quality spectroscopic redshifts of
$B$-band dropouts in the GOODS-N sample distributed over the
range $3.5<z<4.5$, 12 of which belong to the spike. This is a very
strong concentration of galaxies, comparable, for example, to that
discovered at $z\sim 3.09$ by Steidel et al. (1998), and it seems
unlikely that it can be the result of random (Poisson) sampling.
We have generated a large ($10^6$) ensemble of random realizations
of GOODS $B$-band dropouts of the same size as the observed one
using the expected redshift distribution function, and have counted
how often a spike as large as the observed one is realized.  In
principle, one should also take into account the spatial clustering
of $B$-band dropouts in this simulation; in practice, however, the
spatial correlation length, $r_0\sim 4$ comoving Mpc, of these
sources (Lee et al. 2006) is $7\times $ smaller than the radial
distance covered by the redshift bin that we have adopted, $\Delta
r=29$ comoving Mpc, implying that over these scales the
spatial distribution of the galaxies can be approximated as Poisson.

We found no realization in which the number of galaxies in the
redshift bin of GN20 equals or exceeds that of the observed spike,
meaning that the probability to get a similar overdensity by chance
is less than $10^{-6}$.  The simulations show that the average
number of galaxies in the 0.04 redshift bin is $1.31 \pm 1.18$,
implying that the observed spike corresponds to an overdensity of
a factor of 8.4 and significant at the $8.2\sigma$ level.  These
redshift concentrations are frequently found among Lyman-break
galaxies at $z\sim 3$. For example, similar concentrations are found
in every survey field for $U$-band dropouts by Steidel et al. (2003).
Vanzella et al. (2007) also report evidence of concentrations in
the redshift distribution of $B$-band dropouts in the GOODS-South
field.

\begin{figure*}
\centering
\includegraphics[width=16.0cm,angle=0]{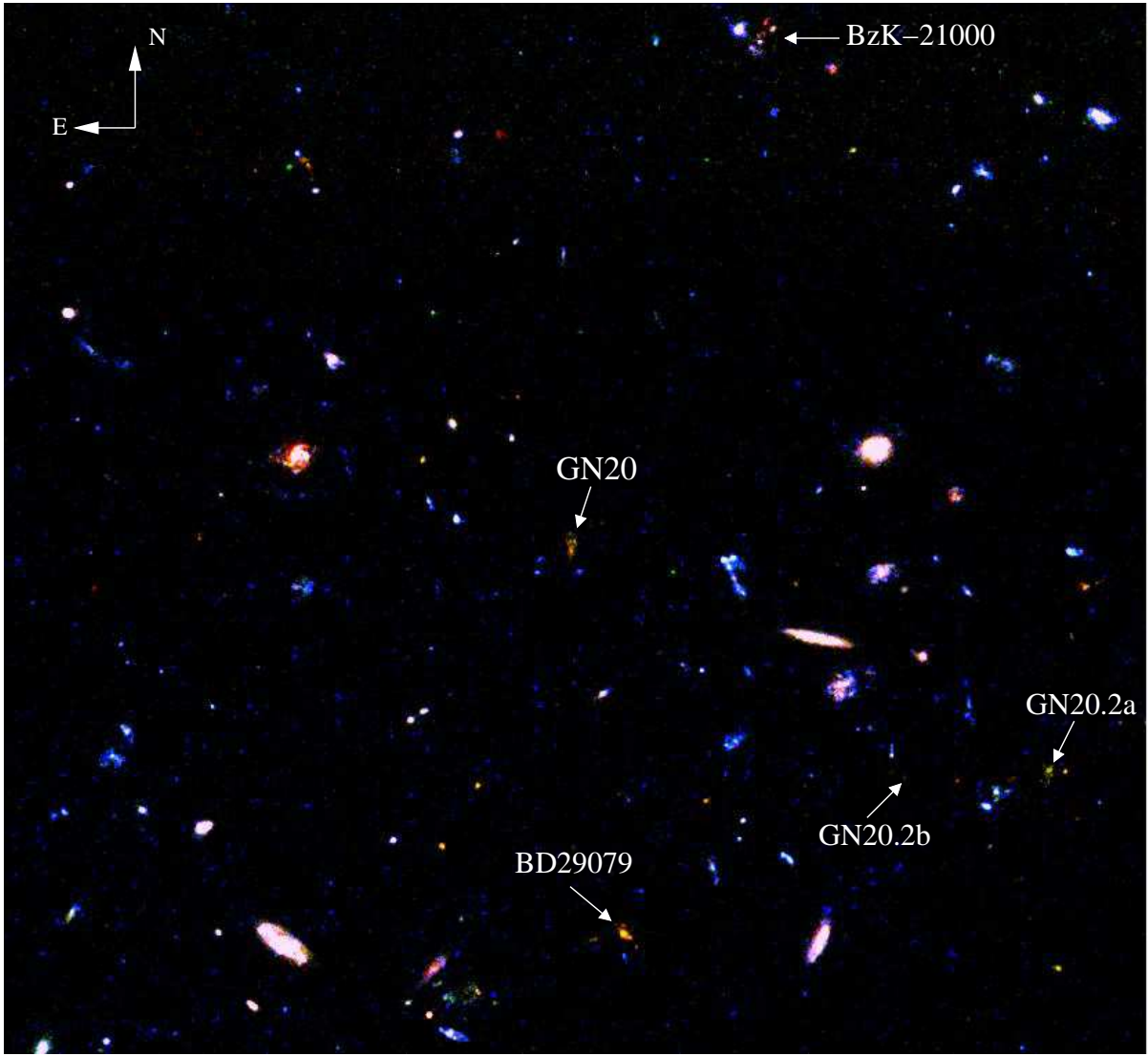}   
\caption{{\it HST}+ACS color image of the field around GN20, derived
from F435W (blue), F775W (green) and F850LP (red). The image size
is $50''$, or 1.77 comoving Mpc at $z=4.05$.  Close-ups of
GN20, GN20.2a, GN20.2b and the luminous "B-drop" are shown in Fig.~\ref{fig:ACS}. The
BzK-21000 galaxy at the top of the image was the original target
of the PdBI observations (Daddi et al. 2008).  Additional dropouts in the
field can be easily recognized from their yellow colors (e.g., see
Fig.~12). 
}
\label{fig:color}
\end{figure*}

To gain additional information about the nature of the GN20 structure,
especially its transverse size, we have studied the distribution
of the angular positions of $B$-band dropouts in the field and
around the GN20 complex. Of the 765 $B$-band dropouts to $z_{\rm
AB}<27$ in GOODS-N, counting all sources regardless of a spectroscopic
identification, 14 are within 25$''$ from GN20 (2.5~Mpc$^2$ comoving;
see Fig.~\ref{fig:color}).  This is an overdensity of a factor of
5.4, the strongest concentration of $B$-band dropout Lyman break
galaxies being present in GOODS-N.  Given that at these angular scales
the angular correlation function of $B$-band dropouts is non
negligible  (Lee et al. 2006), we account for the variance due to
clustering to derive an expected r.m.s fluctuation in the counts
in 25$''$ radius apertures of about 1.8 galaxies.  Therefore, the
excess counts that we see in the 25$''$ radius region surrounding
GN20 correspond to a 5.8$\sigma$ fluctuation.  Taken together with
the analysis of the redshift spike, this leads us to conclude that
the GN20--GN20.2 SMG pair is part of a very strong overdensity  with a
size consistent with a cluster of galaxies, most likely a proto-cluster
given the redshift.

Additional evidence for the presence of a strong overdensity in
this structure comes from Mancini et al. (2008), who use
photometric redshifts to identify galaxy candidates at $z>3.5$ from
an IRAC sample limited to $m_{\rm 4.5\mu m}<23$. They find a total
of 54 candidates in their full GOODS-N IRAC catalogs, which covers
180 arcmin$^2$. 
Of these massive galaxies,
four are within 25$''$ from GN20 (GN20 itself, GN20.2a,
GN20.2b and BD29079; all are shown in
Fig.~\ref{fig:ACS}). This corresponds to an overdensity of a factor
of 18 for this IRAC-selected sample, with a chance probability of
$\sim10^{-4}$  assuming no spatial clustering (Poisson statistics;
0.16 galaxies expected on average and three observed in addition
to GN20).

The possibility that we have uncovered a proto-cluster is reinforced
by the fact that a number of massive galaxies appear to be part of
this structure. With an estimated total baryonic mass of
$\sim2-3\times10^{11}M_\odot$, GN20 is likely one of the most massive
galaxies of this structure. %GN20.2a and GN20.2b together have a similar stellar plus gas mass.  
The presence of over a dozen additional
$B$-band dropouts in this 2~Mpc$^2$ spatial concentration implies
a total baryonic mass reaching $\sim 10^{12}M_{\odot}$ for the overdensity.
If we assume the typical mass-to-light ratio of local massive
clusters of $\sim 50$ (Lin, Mohr \& Stanford 2003), this suggests
a lower limit to the total (dark) mass of the proto-cluster of about
$5\times10^{13}M_\odot$. 

These lines of evidence all support the identification of a large
structure at redshift $z\sim 4$. While its transverse size appears
to be around 2 Mpc (comoving) in diameter, its radial size remains
basically unconstrained given that the errors on the redshift
measurements are already comparable to the transverse size, not
even considering peculiar motions along the radial direction. 
In addition, a significant overdensity at $z=4.05$ seems to be present
extending over the whole GOODS-N field.
This suggests that we are witnessing a proto-cluster of galaxies
which, considering the incompleteness in all the samples we have
been discussing, encompasses a total mass of $\approx 10^{14}M_{\odot}$.

It is interesting to compare the GN20 environment with other
structures previously discovered at high redshift around powerful
steep-spectrum radio galaxies (Kurk et al. 2000; Venemans et al.
2002; 2007; Miley \& de~Breuck 2008 and references therein). The
most direct parallel is with the proto-cluster around TN~J1338-1942
(De Breuck et al. 1999; Venemans et al. 2002) at $z=4.1$. A study
of the overdensity of $B$-band dropout galaxies around this source
has been performed by Miley et al. (2004), reporting factors of
$2.5-5$ excess of Lyman break galaxies (significant at the $3-5\sigma$)
level, depending on the spatial scale.  This is fairly similar to
what is found here for the overdensity around GN20. The field around
TN~J1338-1942 was also observed at 1.2~mm by de Breuck et al. (2004),
finding an overdensity at mm wavelengths as well, although without
spectroscopic confirmations so far.

Also, it is interesting to note that, contrary to the structure described by Chapman et 
al. (2008a) at $z=1.99$ where 5 SMGs are found but without a strong overdensity of UV selected galaxies, 
in this case the two SMGs that we have found at $z=4.05$ (and more SMGs might well be at $z=4.05$ in the GOODS-N field, 
see next Sect.~\ref{sec:phoz}) do appear to coincide with a strong overdensity of Lyman break galaxies as well.

\begin{figure}
\centering
\includegraphics[width=8.8cm,angle=0]{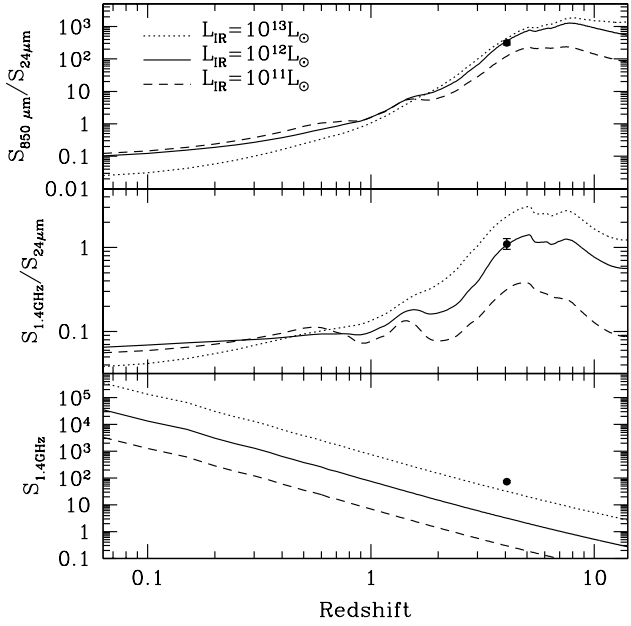}   
\caption{The redshift dependences of several flux density ratios
and for the 1.4~GHz radio flux density expected for CE01 models with 
$L_{\rm IR}=10^{11}L_\odot$ (dashed line),
$L_{\rm IR}=10^{12}L_\odot$ (solid line) and $L_{\rm IR}=10^{13}L_\odot$ (dotted line).
The filled symbols show the values for GN20, which would be detectable
up to $z\sim8$ to
the 5-$\sigma$ limits of $20\mu$Jy in our 1.4~GHz map in GOODS-N (Morrison et al., in preparation).
}
\label{fig:Xreds}
\end{figure}

\begin{deluxetable}{llccccc}%[h]
\tabletypesize{\scriptsize}
\tablecaption{Radio-IR photometric redshifts of SMGs}
\tablewidth{0pt}
\tablehead{
\colhead{ID} &
\colhead{$z_{\rm ORIG}$}&
\colhead{zType}&
\colhead{$z^{\rm radio-IR}_{\rm best}$}&
\colhead{$z^{\rm radio-IR}_{\rm min}$}&
\colhead{$z^{\rm radio-IR}_{\rm max}$}&
\colhead{$\chi^2$}\\
}
\startdata
GN01 & 2.415  & 2 & 2.45 &  2.05 &  2.65  &  8.43 \\  
GN02 & 1.32  & 1 & 1.35 &  0.05 &  2.15  &  13.65 \\  
GN03 & 2.2  & 0 & 1.90 &  0.05 &  2.30  &  7.46 \\  
GN04a & 2.578  & 2 & 2.45 &  2.00 &  3.80  &  0.01 \\  
GN04b & --       & -1 & 3.70 &  2.40 &  9.35  &  0.00 \\  
GN04.2 & 0.851  & 2 & 3.05 &  2.45 &  4.90  &  0.01 \\  
GN05 & 2.21  & 2 & 2.05 &  1.25 &  2.35  &  1.79 \\  
GN06 & 2.00   & 2 & 2.30 &  2.10 &  2.40  &  0.48 \\  
GN07a & --       & -1 & 3.65 &  3.35 &  4.10  &  25.56 \\  
GN07b & 1.998  & 2 & 1.95 &  1.35 &  2.20  &  18.75 \\  
GN09 & 2.9  & 0 & 13.42 &  2.90 &  15.50  &  18.63 \\  
GN10 & 2.2  & 0 & 5.43 &  3.75 &  11.90  &  3.02 \\  
GN11 & 2.3  & 0 & 2.45 &  2.20 &  2.60  &  17.89 \\  
GN12 & 3.10  & 1 & 3.20 &  2.90 &  3.50  &  0.69 \\  
GN13 & 0.475  & 2 & 1.30 &  1.05 &  1.65  &  4.15 \\  
GN15 & 2.74  & 1 & 2.15 &  1.90 &  2.35  &  23.27 \\  
GN16 & 1.68  & 1 & 2.50 &  2.40 &  2.60  &  12.74 \\  
GN17 & 1.73  & 2 & 1.35 &  1.15 &  1.65  &  5.98 \\  
GN18 & 2.8  & 0 & 2.50 &  2.25 &  2.70  &  1.43 \\  
GN19a & 2.484  & 3 & 3.15 &  2.70 &  6.00  &  0.50 \\  
GN19b & 2.484  & 3 & 1.35 &  0.05 &  2.10  &  5.23 \\  
GN20 & 4.055  & 3 & 4.20 &  3.45 &  4.80  &  17.59 \\  
GN20.2a & 4.051  & 3 & 3.70 &  3.35 &  4.05  &  21.27 \\  
%GN20.2b & 4.052$^{\dagger}$  & 3 & 5.05 &  3.50 &  8.00  &  0.17 \\  
GN21 & 2.8  & 0 & 2.95 &  2.60 &  3.60  &  0.25 \\  
GN22 & 2.509  & 2 & 2.95 &  2.60 &  3.50  &  1.21 \\  
GN23 & 2.6  & 0 & 2.95 &  2.55 &  3.35  &  1.48 \\  
GN24 & 2.91  & 1 & 2.15 &  0.10 &  2.70  &  5.16 \\  
GN25 & 1.013  & 2 & 1.15 &  0.35 &  1.40  &  1.48 \\  
GN26 & 1.223  & 3 & 1.82 &  1.50 &  1.95  &  13.94 \\  
GN28 & 1.020  & 2 & 2.95 &  2.55 &  14.60  &  10.38 \\  
GN30 & 1.355  & 2 & 2.15 &  1.45 &  2.45  &  0.37 \\  
GN31 & 0.935  & 2 & 0.90 &  0.05 &  1.30  &  8.86 \\  
GN32 & 1.9  & 0 & 1.75 &  0.05 &  2.35  &  4.84 \\  
GN37 & 3.190  & 2 & 2.90 &  0.10 &  15.50  &  1.89 \\  
LE850.1 & 2.60  & 1 & 2.30 &  2.10 &  2.80  &  10.65 \\  
LE850.4 & 2.60  & 1 & 6.18 &  3.00 &  15.50  &  0.02 \\  
LE850.7 & 1.80  & 1 & 2.20 &  1.95 &  2.45  &  1.18 \\  
LE850.8b & 3.00  & 1 & 2.10 &  1.35 &  2.55  &  2.96 \\  
LE850.8a & 0.974  & 2 & 0.93 &  0.05 &  6.20  &  14.83 \\  
LE850.14b & 2.50  & 1 & 2.35 &  2.05 &  3.05  &  0.56 \\  
LE850.14a & 2.380  & 2 & 2.55 &  1.45 &  4.35  &  0.01 \\  
LE850.18 & 2.690  & 2 & 2.45 &  1.95 &  3.15  &  0.28 \\  
LE850.35 & 3.00  & 1 & 2.40 &  1.85 &  2.80  &  1.11 \\  
Capak	& 4.547 & 3 & 3.97 & 3.00 & 5.20 & 0.22\\
\enddata
\tablecomments{GN galaxies are from P06. LE850 galaxies are from
Egami et al. (2004). The last object is from Capak et al. (2008).
'zType' indicates the origin of stellar photometric redshifts ($z_{\rm ORIG}$): 0 for {\it Spitzer}+IRAC/MIPS photometric
redshifts (P06), 1 for ordinary photometric redshifts from stellar emission, 2 for
spectroscopic redshifts (usually optical/UV redshifts or Spitzer IRS redshifts derived mainly from
detection of PAH features), 3 for CO redshifts. The CO redshifts for
GN19a and GN19b are from Tacconi et al. (2008), the one for GN26
is from Frayer et al. (2008), those for GN20 and GN20.2a are from
this paper. Some of the P06 photometric redshifts listed in P06
were slightly off; here we list the correct values. New spectroscopic
redshifts for three P06 galaxies were measured with IRS in Pope et
al. (2008).  Ranges in $z^{\rm radio-IR}$ are given at the 99\%
confidence level ($\Delta \chi^2=6.63$) following Avni et al. (1976)
for a single interesting parameter in the fit (e.g., redshift).
For the cases of double components we assigned the total submm flux
to each component while computing the photometric redshift. The
photometric redshifts listed here for the GN20 and GN20.2a galaxies
did not use the 3.3mm continuum measurements, for consistency with
the rest of the galaxies. 
}
\label{tab:zphot}
\end{deluxetable}

\section{Radio-IR photometric redshifts for SMGs}
\label{sec:phoz}

An important result of this paper is the success in identifying the
$z\sim4$ redshifts for the GN20 and GN20.2 galaxies based only on
the flux measurements at 24$\mu$m, 850$\mu$m, and 20cm. The ratio
of 20cm to 850$\mu$m flux density is, alone, a powerful redshift
indicator for dusty submm galaxies (Carilli \& Yun 1999; 2000)\footnote{We
applied this redshift indicator to the counterparts of the two SMGs
presented here.  For GN20, we obtain $z=3.74$.  For GN~20.2a, we
obtain $z=1.61$ and for GN~20.2b we obtain $z=4.04$. The
low photometric redshift obtained for GN~20.2a indicates that the
radio emission must be contaminated by an AGN, which substantially
affects the Carilli \& Yun indicator.}, although limited in its
application by its strong dependence on SED temperature. Clearly,
colder galaxies (P06) are intrinsically brighter at 850$\mu$m at
fixed radio flux density and redshift (e.g., P06).  In our approach,
however, we have exploited the 24$\mu$m measurement that lies on
the opposite side of the SED peak. It turns out that for the
variations in the shape of local SEDs (as codified in the CE01
library) the $850\mu$m to 24$\mu$m flux density ratio is fairly
constant at fixed redshift, compared to the redshift excursion of
its value (Fig.~\ref{fig:Xreds}) ---  e.g., this ratio changes by
factors of $\simlt2$ at fixed redshift, while changing the bolometric
luminosity $L_{\rm IR}$ by a factor of 10. However, at fixed $L_{\rm
IR}$ this ratio changes by almost four orders of magnitude when
going from $z=0.5$ to $z=5$ (Fig.~\ref{fig:Xreds}).  Note that we
have not re-normalized the CE01 template models when fitting observed
data from the GN20 and GN20.2 galaxies (Fig.~\ref{fig:Xreds}; bottom
panel).  This information is a powerful complement to the flux
density ratios, as it helps pinpoint the correct intrinsic source
luminosities, thus helping to solve dust temperature degeneracies,
assuming that the local correlations between temperature and
luminosity measured by CE01 and the radio-FIR correlation are not
severely altered in the distant universe. As a result, the photometric
redshift estimates based on IR and radio appear to be robust against
SED temperature/shape variations, as shown by the case of GN20 and
its possibly colder SED (Fig.~\ref{fig:chi2F} and ~\ref{fig:colorS}).

Encouraged by the apparent success of this technique, we explored
in more detail the possibility of deriving accurate radio-IR
photometric redshifts using mid-IR, submm, and radio observations.
We applied this technique to the sample of well-studied SMGs in
GOODS-N (P06) and the Lockman Hole (Scott et al. 2002; Ivison et
al. 2002; Egami et al. 2004), considering only the most likely
counterparts.  Our radio-IR photometric redshifts are determined
as described in Sect.~\ref{sec:chi2IR} for GN20.  In particular,
in order to effectively use the 850$\mu$m information in the fit,
we limit the $S/N$ of the 24$\mu$m and 20cm flux density to a maximum
of 10, artificially increasing the relative error bars if the $S/N$
is higher.  Table~\ref{tab:zphot} summarizes the radio-IR photometric
redshift constraints.

\begin{figure*}
\centering
\includegraphics[width=16.0cm,angle=0]{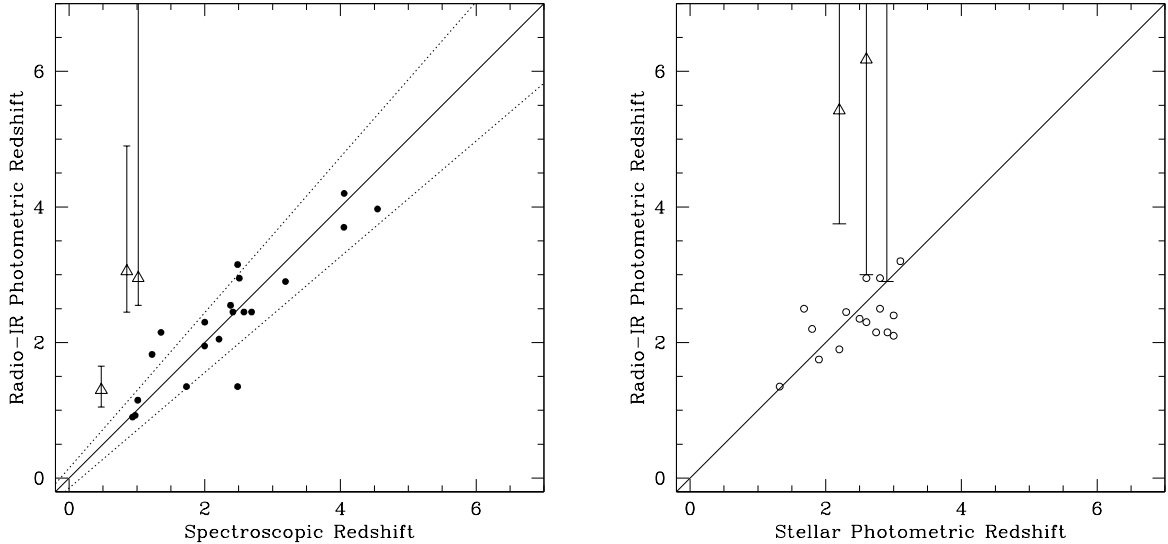}   
\caption{Left panel: radio-IR photometric redshifts for SCUBA
counterparts with known spectroscopic redshifts from P06 for GOODS-N
and Lehmann et al. (2001), Ivison et al. (2005), Chapman et al.(2005)
for Lockman Hole. Empty triangles with error bars (99\% levels) are
GN13, GN04.2 and GN28 from P06.  We suggest that these sources are
not the correct counterparts to the submm emission based on the
discrepant radio-IR photometric and spectroscopic redshifts. Right
panel: the same comparison, but with counterparts having only
photometric redshifts determined from the properties of the stellar
SED, from P06 and Egami et al. (2004).  Triangles with error bars
are sources which are consistent with a $z>4$ solution of the
radio-IR photometric redshifts.
}
\label{fig:xZ}
\end{figure*}

Results of the comparison are in Fig.~\ref{fig:xZ}. In the left
panel, we show SMGs with counterparts having a measured spectroscopic
redshifts. The agreement between radio-IR photometric redshifts and
spectroscopic redshifts are in general fairly good, with a semi
interquartile range of $\Delta z/(1+z)=0.12$. This figure includes
also the Capak et al. (2008) source (assuming
a ratio of 2.8 between the 850$\mu$m and 1.2mm flux densities), but
the result does not change substantially if this one is excluded.

We identify three galaxies as being suspect counterparts to the
submm emission based on their 99\%\, confidence error bars falling
outside the median photometric redshift relationship (Fig.~15).
These three galaxies are from the P06 sample, and they are all
characterized by being at $\simgt6''$ separation or more from the
SCUBA position. Given that P06 finds a median separation of $\sim3''$
for secure counterparts, with 80\% being at $<4.5''$ \footnote{Dannerbauer
et al. (2004) also finds a median separation of 3$''$ for the radio
counterparts of MAMBO sources, with a similar percentage of radio
counterparts being separated by less than 4.5$''$.}, the large
offsets alone suggest that the counterpart identification might
have actually failed for these objects. Only three out of 43 targets
are identified as such, and two of these were already listed as
tentative counterparts, confirming the high quality of counterpart
identification in the original work.  We discuss these three objects
here in some detail:\\ $\bullet$ GN13: the counterpart in P06 is a
galaxy at $z_{\rm spec}=0.475$ lying at $7''$ from the submm position.
The identification was judged to be secure, on the basis of the low
probability of a chance association between a radio and mid-IR
bright source relative to the submm position.  Its 850$\mu$m flux
density of 1.9~mJy, despite being quite faint intrinsically, is
about six times brighter than expected on the basis of both the
24$\mu$m and 20cm radio flux densities for its redshift, assuming
the CE01 templates.  We inspected the other galaxies present in the
submm beam, and found no other convincing counterpart. This SMG
could be at quite high redshift.\\ $\bullet$ GN04.2: the counterpart
listed by P06 is at $z_{\rm spec}=0.851$ and $5.7''$ from the submm
position. The 24$\mu$m and radio flux densities are both too faint
to be consistent with the 850$\mu$m flux density of 2.7~mJy at this
redshift, assuming the CE01 templates.  Furthermore, this source
is no longer detected in the latest and deepest 20cm reduction
(Morrison et al., in preparation); the P06 results were based on a
$3\sigma$ detection at 20cm. No other 24$\mu$m detected galaxy is
present in the beam, which may imply that the real counterpart is
at fairly high redshift.\\ $\bullet$ GN28: the counterpart listed
by P06 is at $z_{\rm spec}=1.020$ and at $5.8''$ from the submm
position. There is a strong inconsistency between the faint 24$\mu$m
flux density of this galaxy (20.5$\mu$Jy) and its 850$\mu$m emission
(1.7~mJy). The 850$\mu$m to 24$\mu$m ratio is two orders of magnitude
larger than expected for starforming galaxies at $z=1$
(Fig.~\ref{fig:Xreds}); and would be even more discrepant for an
AGN SED.  The 20cm radio position is close to the radio lobes of a
bright radio galaxy at a similar redshift, and thus may be severely
contaminated.  No other 24$\mu$m detected galaxy is present in the
beam, which may also imply that the real counterpart is at fairly
high redshift.  

When excluding these three likely incorrect
counterparts, we find $\sigma(\Delta z/(1+z))=0.15$ and a
semi-interquartile range of $\Delta z/(1+z)=0.08$.

The right panel of Fig.~\ref{fig:xZ} shows the comparison of the
radio-IR photometric redshifts with those counterparts of SMGs from
the same works that have only photometric redshift estimates based
largely on the stellar emission (plus the IRAC to MIPS colors in a
few cases from P06).  Again, the agreement is quite good, with a
similar $\sigma(\Delta z/(1+z))=0.14$. We find only two cases of
widely discrepant estimates, discussed here in some detail:\\
$\bullet$ GN10: P06 assign a $z=2.2$ photometric redshift, while
we find $z>3.7$ at the 99\% confidence level, in good agreement
with Wang et al. (2007; 2008) and Dannerbauer et al. (2008).  The
discrepancy is likely due to the photometric redshift of P06 being
too low, while the counterpart {identification appears correct and
agrees with Wang et al. (2007; 2008) and Dannerbauer et al. (2008)}.\\
$\bullet$ GN16: The P06 identification appears correct, and the
discrepancy is due to the low photometric redshift ($z_{\rm
phot}=1.68$) estimate of P06.

\begin{deluxetable*}{lcccccc}%[h]
\tabletypesize{\scriptsize}
\tablecaption{Properties of the $z>3.5$ starburst candidates selected using their high radio 1.4~GHz
to 24$\mu\lowercase{\rm m}$ flux density ratios
}
\tablewidth{0pt}
\tablehead{
\colhead{ID} &
\colhead{RA(J2000)}&
\colhead{DEC(J2000)}&
\colhead{$S_{\rm 1.4~GHz}$}&
\colhead{$S_{\rm 24\mu m}$}&
\colhead{$S_{\rm 8.0\mu m}$}&
\colhead{$S_{\rm 4.5\mu m}$}
\\
\colhead{}&
\colhead{}&
\colhead{}&
\colhead{($\mu$Jy)}&
\colhead{($\mu$Jy)}&
\colhead{($\mu$Jy)}&
\colhead{($\mu$Jy)}
}
\startdata
VLA~512 &     12:36:58.49 & 62:09:31.8 & 47.3 $\pm$ 11.5 & 36.8 $\pm$ 5.0 & $9.7\pm1.0$ & $4.0\pm0.4$\\ 
VLA~812 &     12:35:53.24 & 62:13:37.5 & 45.6 $\pm$  5.1 & 33.0 $\pm$ 6.2 & $23.1\pm2.3$ & $14.6\pm1.5$\\
VLA~966 &     12:37:01.52 & 62:20:24.6 & 64.2 $\pm$  5.3 & 49.7 $\pm$ 3.6 & $22.9\pm2.3$  & $14.2\pm1.4$\\
\enddata
%\tablecomments{
%}
\label{tab:radio}
\end{deluxetable*}

We have also applied the radio-IR photometric redshift technique
to the CO-detected SMG galaxy reported by Capak et al. (2008) and Schinnerer et al. (2008)
to be at $z=4.547$.
The resulting photometric redshift is $z=4.0$ and the 99\% confidence
range is $3.0<z<5.2$, supporting the $z=4.5$ redshift
identification of this SMG despite the large offset ($\simgt1''$;
$\simgt 10$~kpc proper) between its Ly$\alpha$ emitting optical
counterpart and the radio emission.  
%Detection of CO emission would
%be required to confirm the identification, but our results imply
%that, in any case, this SMG is at very high redshift.

In conclusion, we find that radio-IR photometric redshifts can be
used as a powerful and independent complement to photometric redshifts
derived from stellar emission, and could greatly improve the SMG
counterpart identification process. Some differences and evolutionary
trends are seen when studying the SEDs of SMGs in detail (e.g.,
Pope et al. 2006; 2008).  However, the fact that direct fitting of
locally-derived CE01 templates yields reasonable photometric redshifts
without any significant bias with respect to redshift suggests that
ULIRGs SEDs have not dramatically evolved from $z=0$ to $z=4$.

\section{The selection and space density of the highest redshift starburst galaxies}
\label{sec:z4}

We emphasize that not all radio faint (e.g., $S_{\rm 1.4~GHz}<30\mu$Jy)
counterparts to SMGs are likely to be at $z\simgt 4$.  For example,
four SMGs (GN03, GN11, GN14 (HDF~850-1), GN18) have even fainter
radio flux densities than GN10 and have similar properties to GN10
at other wavelengths, e.g. they are undetected in ACS imaging and
have red IRAC SEDs.  Dannerbauer et al. (2008) suggest they could
lie at $z\simgt4$. In contrast, the radio-IR photometric redshifts
suggest that GN03, GN11 and GN18 are at $2<z<3$, which is still
consistent with their IRAC SEDs. This is because these galaxies
have quite bright $24\mu$m flux densities, much larger than the
radio ones.  Besides GN20, GN20.2a and GN20.2b, only four additional objects
from P06 are identified by the photometric redshift technique to
possibly be at $z\simgt4$. With various degrees of confidence, they
are GN10, a very solid $z\simgt4$ source also identified by Wang
et al. (2007; 2008) and Dannerbauer et al. (2008), and GN13, GN28
and GN04.2 (where reliable counterparts are, however lacking, thus
making these less convincing candidates). Another well known
$z\simgt4$ candidate from GOODS-N is the galaxy HDF850.1 (Dunlop
et al. 2004), listed as GN14 in P06 for which we have not obtained
a radio-IR photometric redshift due to the lack of reliable radio
and 24$\mu$m data. This would make a total of eight plausible $z>3.5$
SMGs in GOODS-N in a total area of roughly $50-100$ arcmin$^2$
covered with SCUBA (with a highly inhomogeneous coverage; see, e.g.,
Pope et al. 2005), or roughly $200-300$ galaxies deg$^{-2}$.  Using
the volume in $3.5<z<6$, this corresponds to a space density of
$\approx 10^{-5}$~Mpc$^{-3}$, which compares fairly well with the
expectations for the very high redshift formation of the oldest
early type galaxies seen at $z\sim1.5-2.5$ (see Sect.~1).  With
typical SFRs in excess of thousand $M_\odot$~yr$^{-1}$, such a
population contributes an SFRD of order of
$2\times10^{-2}$~$M_\odot$~yr$^{-1}$~Mpc$^{-3}$, already
comparable to the global contribution of Lyman break galaxies at
these redshifts (Giavalisco et al. 2004b).

\subsection{A radio/24$\mu$m selection criterion for high-$z$ starburst galaxies}

For the most reliable $z\simgt4$ starbursts (GN10, GN20, GN20.2a)
we notice that all of these have very high 20cm to 24$\mu$m flux
density ratios, greater than about 1 (Fig.~\ref{fig:Xreds}).  GN20.2b
itself satisfies such a criterion and might also be  at $z=4.05$.
This suggests that a $S_{\rm 1.4~GHz}\simgt S_{\rm 24\mu m}$ criterion
could be useful at singling out the very high redshift tail of SMGs.
Applying this criterion to the P06 sample we recover the SMGs listed
above, plus GN12, GN16 and GN19a, all sources at $2.5<z\simlt3.2$,
a distribution clearly skewed to higher redshifts than typical SMGs.
The $z=4.547$ galaxy from Capak et al. (2008) also satisfies this
criterion.

In principle, this simple criterion could be applied regardless of
a submm detection, and could avoid the inherent bias toward
intrinsically cold sources. Of course, care should be taken and a
careful analysis of all multiwavelength information should be
performed on targets selected in this way, as radio emission from
radio-loud AGN would bias the radio to 24$\mu$m flux density ratio.
Mid-IR as well can be boosted by AGNs, which could make genuine
$z>4$ objects fail this criterion. More importantly, given the
relatively small dynamic range expected over redshift for the 20cm
to 24$\mu$m flux density ratios in galaxies (Fig.~\ref{fig:Xreds}),
scatter due to SED variations can imply a substantial contamination
from lower redshift sources.

We have performed a preliminary test of this criterion with a sample
of radio selected galaxies with $S_{\rm 1.4~GHz}>30\mu$Jy in GOODS-N
(Morrison et al., in preparation), applying a simple but very
conservative approach.  By requiring $S_{\rm 1.4~GHz}>0.8\times
S_{\rm 24\mu m}$, 63 radio sources are retained.  Only half of these
(34) are undetected in the ACS F435W band ($S/N<2$), a necessary
condition for being at $z\simgt4$. Of these, 16 are also consistent
with a red IRAC SED, with a peak beyond $7\mu$m, as expected for
$z\simgt 4$\footnote{The Capak et al.  (2008) galaxy does not satisfy
this requirement. The authors suggest that line emission contaminate
the IRAC photometry for this source. A possible alternative is that
the IRAC flux detected is actually from a foreground, lower redshift
galaxy. This would imply that the SMG is very faint even at the
IRAC bands, perhaps similar to GN10.}. We further require relatively
blue {\it Spitzer}+MIPS to IRAC flux density ratios: $S_{\rm
24\mu m}/S_{\rm 4.5\mu m}<13$ and $S_{\rm 24\mu m}/S_{\rm
8.0\mu m}<5$, as observed in the bona-fide $z\simgt4$ sample (GN20,
GN20.2a, GN20.2b, GN10), leaving us with 11 galaxies. This excludes from
the sample the $z=4.424$ radio source from Waddington et al. (1999),
where the near-IR to mid-IR spectral range is likely dominated by
a powerful AGN.  Finally, we exclude all sources detected in the X-ray catalog
of Alexander et al. (2003)
which, if at $z\simgt 4$, must be AGN dominated.  This leaves us
with seven sources, including the three  reliable $z>4$ galaxies
already known (GN20, GN20.2a, GN10) plus GN20.2b.  We consider the
properties of the three new candidates in detail:\\ $\bullet$ VLA~512:
this object is undetected in a relatively shallow region of the
SCUBA supermap. However, some signal is found at its position with
$S_{\rm 1.1mm}=3.2$mJy ($S/N=3.4$) in an AzTEC map of GOODS-N
(G.~Wilson, private communication).  This source has no counterpart
at any ACS band, and is a good high redshift starburst candidate.
\\ $\bullet$ VLA~812: this galaxy is very close to the edge of the
SCUBA supermap, where the noise is fairly high. However, this source
is listed in Chapman et al. (2005) as a SCUBA detection in photometry
mode ($S_{\rm 850\mu m}=8.8\pm2.1$mJy), with a redshift of $z=2.098$ 
(the source is also detected by Perera et al. 2008 and Greve et al. 2008
at 1.1mm and 1.2mm with MAMBO and AzTEC, respectively).
We find that a relatively bright $z=2.098$ galaxy is present in the
area, but 2$''$ away from the IRAC and radio position, which
corresponds instead to a faint galaxy with ACS photometry of $i_{\rm
AB}=27.6$ and $z_{\rm AB}=27.3$, undetected at $B$- and
$v$-band. We suggest that the Chapman et al. (2005) redshift
identification does not correspond to the actual counterpart of the
radio and submm emission. This radio source is a reliable candidate
for a very high redshift starburst ($z\simgt4$).\\ $\bullet$ VLA~966:
we place a 3$\sigma$ upper limit of $S_{\rm 850\mu m}<8.1$mJy from
the SCUBA supermap at the position of this source. Its optical
counterpart is a $B$-band dropout with $z=26.6$ and red IRAC SED,
confirming that this is a reliable $z\sim4$ candidate.  The VLA
radio flux density of this source is comparable to that of GN20.
The lack of submm emission could signal a warmer SED (see, e.g., Chapman et al. 2008b).

We notice that, due to the very conservative requirements applied, we finally retained only
3 galaxies of the 63 selected down to 30$\mu$Jy in GOODS-N
with our radio/24$\mu$m criterion as bona-fide $z\simgt3.5$ starburst galaxy candidates. 
We believe that we efficiently rejected in this way all contamination from 
AGN dominated radio galaxies. Several of the discarded galaxies are though most likely to 
be starbursts at $2.5\simlt z\simlt3.5$, still very interesting objects but beyond the scope
of the simple exercise described in this Section where we focus on the $z>3.5$ redshift tail. 
We are likely also missing several other genuine 
$z\simgt3.5$ starburst, due to the very strict requirements that we have applied. 
E.g., that of a distinct peak in IRAC beyond
$7\mu$m that can be weak or washed out due to the noise to the levels of our data; or that of X-ray non detections,
powerful $z\simgt 3.5$ starbursts can well contain some unobscured AGN sometimes that doesn't necessarily affect
much the radio emisson. Given the promising results of this attempt, we will defer to future work 
an investigation of the properties of the full sample of galaxies selected with this  radio/24$\mu$m criterion.

Although we by no means claim to have selected a complete sample,
the existence of these three additional $z>3.5$ starburst candidates
reinforces our conclusion that a substantial population of vigorous
star-forming galaxies was already present at early epochs.

\subsection{Predictions for future {\it Herschel} surveys}

We now consider expectations for forthcoming {\it Herschel} surveys
to detect very high redshift starbursts. {\it Herschel} will obtain
FIR imaging with an unprecedented sensitivity at 70$\mu$m, 100$\mu$m
and 160$\mu$m with PACS (Poglitsch et al. 2006) and at 250$\mu$m,
350$\mu$m and 520$\mu$m with SPIRE (Griffin et al. 2007) .

We computed the expected flux densities of GN20 in the {\it Herschel}
bands, based on various assumptions.  Using the best fitting $z=4.055$ {\em unscaled} CE01
model shown in Fig.~\ref{fig:chi2F}, we predict flux densities of
about $40-50$~mJy in the SPIRE bands, 24~mJy at 160$\mu$m, 4.8~mJy
at 100$\mu$m and 1.8~mJy at 70$\mu$m. The latter value is comparable,
but still consistent, with the {\it Spitzer} 70$\mu$m upper limit.
If these predictions are correct, galaxies similar to GN20 should
be very easily detectable with {\it Herschel}, e.g., over the full
70~deg$^2$ survey of the Hermes guaranteed time consortium\footnote{\tt
http://astronomy.sussex.ac.uk/$\sim$sjo/Hermes/.}, which could yield
hundreds $z>4$ hyperluminous infrared galaxies.  Even if
we used a colder template following Fig.~\ref{fig:colorS}, normalized
conservatively at a 1$\sigma$ lower 850$\mu$m flux density of 17~mJy,
we would still predict flux densities of about 30~mJy at 350$\mu$m
and 520$\mu$m, which is still above the expected flux limit of the
70~deg$^2$ Hermes survey.

This suggests that, while strongly limited by confusion to quite
bright flux densities ($\sim 15$~mJy; Le~Borgne et al. 2008), the
{\it Herschel} SPIRE observations will efficiently detect a significant
tail of very high redshift ($z>4$) starbursts.  Due to the relation
between redshift and 850$\mu$m flux density (Ivison et al. 2002; Pope et al 2006),
{\it Herschel} might detect the bulk of the most active merger-powered
galaxies in the distant universe, and possibly accounting for an
important fraction of the SFRD in the early universe.  Clearly, a
real challenge will be to recognize such very high redshift galaxies
in the {\it Herschel}-selected samples, which will require deep
multi-wavelength information.

\section{Summary and Conclusions}
\label{sec:conclusion}

The serendipitous detection of CO emission lines at $z>4$ from deep
PdBI observations allowed us a first investigation of the nature
and properties of vigorous star forming galaxies at these
redshifts. The main results of this paper can be summarized as
follows:

\begin{itemize}

\item CO emission lines are detected at the position
of two bright SMGs in the GOODS-N field: GN20 and GN20.2a. A tentative
CO signal is seen for an additional radio galaxy in the field, named
GN20.2b.

\item We identify the lines of GN20 and GN20.2a as CO[4-3] at $z=4.05$. A concordant
identification is obtained when comparing optical photometric
redshifts, ACS dropout color selection, radio-IR photometric redshifts,
and optical spectroscopy from Keck.

\item Dust continuum at 3.3mm (0.65mm rest frame) is significantly detected for GN20, 
the first such a measurement in a distant starburst galaxy.

\item We show that reliable radio-IR photometric redshifts (to $\Delta
z\sim 0.5$) can be derived for these $z=4.05$ galaxies using only the
photometry at 24$\mu$m, 850$\mu$m, 3.3mm and radio 20cm and CE01
models. This implies that
local models can be used to provide a reasonable description of the $z=4.05$
galaxies. 

\item However, evidence is also found for these $z\sim4$
SEDs being somewhat colder than those
of local galaxies of the same luminosities, confirming the results of P06 
for lower redshift objects. 
In addition, the strong radio flux density of GN20.2a
is taken as evidence for the presence of an AGN in this galaxy.

\item The IR to radio SEDs suggest very high IR luminosities of 2.9 and 1.6 $\times10^{13}L_\odot$
for GN20 and GN20.2, respectively. The CO emission lines imply molecular gas masses of 5 and 3 $\times10^{10}M_\odot$
for GN20 and GN20.2, respectively. For GN20 we could derive an estimate of the dynamical mass of order of
2$\times10^{10}M_\odot$, implying molecular gas fraction of order of 20\% 
(or larger if the CO[4-3] transition is not thermalized).

\item The FIR to CO properties and SSFRs of these $z\simgt4$ SMGs are similar to
SMGs
at lower redshift, $1.5<z<3$.

\item Radio and submm stacking of $B$-band dropout galaxies, together 
with the SED fitting of the ACS to IRAC SEDs, is used to
show that typical $z \sim 4$ starforming galaxies have much lower SSFRs
than SMGs at the same redshift. This suggests that the
stellar mass-SFR correlation does not evolve much further in normalization beyond
$z\sim2$.

\item The GN20 and GN20.2 SMGs appear to live at the center of a
strong sky overdensity of galaxies (with an
apparent size of $1-2$ comoving Mpc), and a very significant redshift spike
is seen over the full GOODS-N area, consistent with a proto-cluster.
It is tantalizing to speculate that this overdense environment might be 
playing a role in triggering such extreme star formation activity
(see, e.g., Elbaz et al. 2007; Chapman et al. 2008b).

\item Applying the newly defined radio-IR photometric redshift
technique to the full sample of SMGs selected in the GOODS-N and
Lockman~Hole fields, we find that this provided estimates with a
typical accuracy of $\Delta z/(1+z)\sim0.15$. This is starting to
approach the accuracy obtainable from optical photometric redshifts
($\Delta z/(1+z)\sim0.06$; e.g., Brodwin et al. 2006).

\item Using this photometric redshift technique, we identify a small
sample of galaxies in GOODS-N whose counterpart identification had
likely failed in previous works. Lacking an obvious counterpart within
the SCUBA beam, these SMGs could be at quite high redshift.

\item A total of up to eight SMGs in GOODS-N (including GN20, the two counterparts of
GN20.2, GN10, HDF850.1, GN13, GN28, GN04.2) are found consistent with being
at $z\simgt4$. This implies a space density of similar sources around
$10^{-5}$~Mpc$^{-3}$ and a sizeable contribution to the cosmic SFRD at
$z\sim4$ of order of $10^{-2}M_\odot$~yr$^{-1}$~Mpc$^{-3}$.  This is similar to
that of UV-selected galaxies at the same redshifts.

\item We find that the GN20, GN20.2a and GN20.2b galaxies and the most secure
$z>4$ SMG candidates satisfy the simple criterion $S_{1.4~GHz}\simgt
S_{24\mu m}$. By applying this criterion, with additional conservative
constraints from optical, near-IR and mid-IR photometry, to a sample
of VLA 20cm selected galaxies in GOODS-N, we identify three additional
galaxies that appear to be reliable $z>4$ starburst galaxy
candidates. A posteriori, two of the three are found to have a
detection at submm/mm wavelengths as well.

\item Given the expected FIR fluxes of these $z>4$ galaxies, future
surveys with Herschel should allow us to characterize the properties of
this population in detail.

\end{itemize}

\acknowledgements Based on observations carried out with the IRAM
Plateau de Bure Interferometer. IRAM is supported by INSU/CNRS
(France), MPG (Germany) and IGN (Spain). We are grateful to Grant
Wilson for sharing with us unpublished information from the GOODS-N
AzTEC supermap.  We thank Arancha Castro-Carrizo for 
assistance with the reduction of the IRAM Plateau de Bure Interferometer
data, and Ranga-Ram Chary for discussions.  Comments by an anonymous referee helped improving the paper.
We acknowledge the use of GILDAS software ({\tt
http://www.iram.fr/IRAMFR/GILDAS}).  ED, CM and DE acknowledge support
from the French ANR grant numbers ANR-07-BLAN-0228 and ANR-08-JCJC-0008. 
The work of DS
was carried out at Jet Propulsion Laboratory, California Institute of
Technology, under a contract with NASA.  Support for this work was
provided by NASA, Contract Number 1224666 issued by the JPL, Caltech,
under NASA contract 1407.

\end{document}